\documentclass[letterpaper,onecolumn,10pt]{quantumarticle} % use a4paper instead of letterpaper if you prefer
\pdfoutput=1

\usepackage[utf8]{inputenc}
\usepackage[T1]{fontenc}
\usepackage[english]{babel}
\usepackage{amsmath,amssymb,amsfonts}
\usepackage{mathtools}
\usepackage[numbers,sort&compress]{natbib}
\usepackage{graphicx}
\usepackage{xcolor}
\usepackage{booktabs}
\usepackage{longtable}
\usepackage{bookmark}
\usepackage{xurl}
\usepackage{microtype}
\usepackage{subcaption}
\usepackage{tabularx}
\usepackage{makecell}
\usepackage{siunitx}
\usepackage{tikz}
\usetikzlibrary{arrows.meta,plotmarks,positioning,calc,matrix}
\usepackage{pgfplots}
\usepgfplotslibrary{dateplot}
\pgfplotsset{compat=1.18}
\DeclareSIUnit{\hash}{hash}
\DeclareSIUnit{\USD}{\text{\$}}
\newcolumntype{Y}{>{\raggedleft\arraybackslash}X}
\newcolumntype{Z}{>{\raggedright\arraybackslash}X}
\newif\ifhasquantikz
\IfFileExists{quantikz.sty}{%
  \usepackage{quantikz}%
  \hasquantikztrue
}{%
  \hasquantikzfalse
}
\usepackage{xspace}
\newcommand{\SHA}{\mbox{SHA-256}\xspace}
\newcommand{\dSHA}{\mbox{double-SHA-256}\xspace}
\newcommand{\RIPEMD}{\mbox{RIPEMD-160}\xspace}
\newcommand{\PtwoPKH}{\mbox{P2PKH}\xspace}
\newcommand{\secref}[1]{Sec.~\ref{#1}}
\newcommand{\eqnref}[1]{Eq.~\eqref{#1}}
\newcommand{\Tdiff}{T_{\mathrm{diff}}(n)}
\newcommand{\Toracle}{T_{\mathrm{oracle}}}
\newcommand{\Ttot}{T_{\mathrm{tot}}}

\title{Kardashev scale Quantum Computing for Bitcoin Mining}

\author{Pierre-Luc Dallaire-Demers}
\thanks{Research funded and supported by BTQ Technologies.}
\affiliation{Pauli Group}
\orcid{0000-0002-9316-5597}
\email{pierre-luc@pauli.group}

\author{BTQ Technologies Team}
\affiliation{BTQ Technologies}

\date{}

\begin{document}

\maketitle

\begin{abstract}
Bitcoin already faces a quantum threat through Shor attacks on elliptic-curve signatures.
This paper isolates the other component that public discussion often conflates with it: mining.
Grover's algorithm halves the exponent of brute-force search, promising a quadratic edge to any quantum miner of Bitcoin.
Exactly how large that edge grows depends on fault-tolerant hardware. No prior study has costed that hardware end to end.
We build an open-source estimator that sweeps the full attack surface: reversible oracles for double-\SHA mining and \RIPEMD-based address preimages, surface-code factory sizing, fleet logistics under Nakamoto-consensus timing, and Kardashev-scale energy accounting.
A parametric sweep over difficulty bits $b$, runtime caps, and target success probabilities reveals a sharp transition.
At the most favourable partial-preimage setting ($b=32$, $2^{224}$ marked states), a superconducting surface-code fleet still requires $\sim\!10^{8}$ physical qubits and $\sim\!10^{4}\,$MW. That load is comparable to a large national grid.
Tightening to Bitcoin's January~2025 mainnet difficulty ($b\approx79$) explodes the bill to $\sim\!10^{23}$ qubits and $\sim\!10^{25}\,$W, approaching the Kardashev Type~II threshold.
These numbers settle a narrower question than ``Is Bitcoin quantum-secure?''
Once Grover mining is lifted from asymptotic query counts to fault-tolerant physical cost, practical quantum mining collapses under oracle, distillation, and fleet overhead.
To push mining into non-trivial consensus effects, one must invoke astronomical quantum fleets operating at energy scales that lie far above present-day civilization.
\end{abstract}

\tableofcontents
\clearpage
\section{Introduction}

Fifteen gigawatts power the Bitcoin network. That exceeds the electricity use of many
nation-states. All of it buys one guarantee: no adversary finds a
valid block header faster than brute force
permits~\cite{Nakamoto2008,CBECI2025}.
Grover's algorithm threatens that guarantee.
A quadratic speedup halves the exponent of proof-of-work search; a quantum
miner armed with error-corrected Grover oracles could, in principle, locate
a winning nonce with the square root of the classical
work~\cite{Grover1997}.
``Can quantum computers break Bitcoin mining?''
The question echoes across the blockchain community, yet the literature
offers no definitive answer. Prior analyses do one of two things.
They either halt at asymptotics, where ``Grover gives $O(\sqrt{N})$,''
or they bury the verdict inside surface-code ledgers that practitioners cannot
extract~\cite{Amy2017-SHA,Gheorghiu2019-Mosca,Babbush2021}.
That ambiguity matters because Bitcoin's quantum story already splits in two.
Shor's algorithm threatens the signature layer on a nearer-term front~\cite{DallaireDemers2025-ECDLP}, while public discussion often treats the mining layer as vulnerable for the wrong reason: it imports Grover's asymptotics and skips the physical bill.
This paper addresses that second question.

Three compounding overheads hide behind that clean asymptotic promise.
A reversible double-SHA-256 oracle demands
${\sim}\,3\times10^{5}$ non-Clifford ($T$) gates per Grover iteration,
inflating the na\"ive query count by a factor of
${\sim}\,2^{38}$~\cite{Amy2017-SHA}.
Each $T$ gate in turn requires a magic-state distillation factory that
occupies thousands of physical qubits on the surface
code~\cite{Litinski2019}.
Nakamoto's ten-minute block window then caps the depth of any single
search and forces the adversary to field exponentially many independent
machines. Each machine is too slow on its own to reach the target success
probability~\cite{Sattath2020}.
Oracle cost, distillation tax, fleet multiplication: together these three
factors transform a quadratic speedup into a resource bill that no single
study has totalled.

This paper closes the loop.
We trace the full path: reversible oracle design, surface-code factory
sizing, fleet logistics, and wall-plug power. We then sweep difficulty bits,
runtime caps, and success thresholds across three hardware architectures:
superconducting, trapped-ion, and neutral-atom.
The estimator converts fleet qubit counts into watts and maps each result
onto Kardashev's civilization ladder, where Type~I marks planetary power
(${\sim}\,10^{16}\,$W), Type~II stellar (${\sim}\,10^{26}\,$W), and
Type~III galactic (${\sim}\,10^{36}\,$W)~\cite{Kardashev1964}.

The verdict falls hard against the quantum miner.
At the most favourable partial-preimage setting, namely difficulty $b=32$ with
$2^{224}$ marked states, a superconducting fleet still consumes
${\sim}\,10^{4}\,$MW across ${\sim}\,10^{8}$ physical qubits, a load that
rivals a large national grid.
Tighten to Bitcoin's January~2025 mainnet difficulty ($b\approx79$) and
the bill explodes: ${\sim}\,10^{23}$ qubits drawing
${\sim}\,10^{25}\,$W, which approaches the Kardashev Type~II threshold.
Grover's quadratic speedup does not survive the multiplicative overhead of
fault tolerance at Bitcoin-relevant scales.
The moral is narrow and sharp.
This paper does not revisit the signature-layer Shor threat~\cite{AggarwalEtAl2018QuantumAttacksBitcoin}. It shows that the mining component fails for the opposite reason. That is the part often cited in public discussions of ``quantum Bitcoin mining.''
To produce non-trivial effects on Nakamoto consensus, a Grover miner must command astronomical fleets operating at energy scales that only much higher Kardashev civilizations could plausibly support.
Figures~\ref{fig:fleet-heatmap} and~\ref{fig:fleet-tradeoff} show the baseline machine in fleet units: the favourable partial-preimage corner already demands industrial-scale hardware, and tighter runtime caps force explosive growth in machine count.
Figure~\ref{fig:kardashev-energy-budget} turns that fleet into a power budget and pushes the mainnet point toward Type~II.
Figures~\ref{fig:energy-scale-ladder}--\ref{fig:high-energy-tradeoff} then test the science-fiction escape hatch.
The ladder buys speed, but the high-energy heatmaps and trade-off curves still demand objects that read less like computers than like engineered astrophysical bodies.

The remainder of the paper assembles this verdict piece by piece:
background on Grover search and surface-code accounting
(\secref{sec:related-work}), the reversible oracle ledgers that price each
hash call (\secref{sec:logical-grover}), the surface-code fleet mapping
that converts logical gates into physical qubits
(\secref{sec:physical-footprint}), the Kardashev-scale energy analysis
(\secref{sec:energy-kardashev}), and the implications for Nakamoto
consensus (\secref{sec:discussion}).
The route follows the debunking trail in order: oracle cost, distillation tax, fleet multiplication, then power.
All estimator code, test suites, and sweep data that reproduce every table
and figure appear in a public
repository.\footnote{\url{https://github.com/Pauli-Group/kardashev}}

\section{Background and related work}
\label{sec:related-work}

\subsection{Grover's Algorithm and Cryptographic Search}
Grover's algorithm~\cite{Grover1997} locates a marked item in an unstructured set of size $N$ with $O(\sqrt{N})$ queries. That is a quadratic cut against brute force.
In cryptography, $n$-bit preimage resistance drops to roughly $n/2$ bits under quantum queries.
For Bitcoin's proof-of-work, the predicate reads ``$\mathrm{SHA}\text{-}256(\cdot)$ falls below target,'' so the oracle implements the \SHA compression function~\cite{FIPS180-4,Nakamoto2008}.
The asymptotics look simple; the \emph{physical} bill on a fault-tolerant device does not: error-correction overhead, clock rate, and oracle cost drive the total.
For collision resistance, quantum algorithms achieve \(O(N^{1/3})\) complexity~\cite{Brassard1997Collision}, but a collision search does not help Bitcoin's thresholded proof-of-work predicate or single-target preimage on \PtwoPKH, so we focus on preimage oracles.

\subsection{Fault-Tolerance, Surface Code, and Space--Time Overheads}
Large Grover searches demand full fault tolerance.
The surface code leads on threshold and locality, yet it exacts heavy qubit and time costs, with a premium on non-Clifford (\emph{T}) gates that require magic-state distillation.
Litinski's framework~\cite{Litinski2019} exposes a space--time tradeoff: slow runs with few qubits; fast runs with many.
At a physical error rate $10^{-4}$ and a $1\,\mu\mathrm{s}$ code cycle, a 100-qubit job with $T$-count $10^8$ and $T$-depth $10^6$ finishes in $\sim4$ hours with $\sim5.5\times10^4$ physical qubits, in $\sim22$ minutes with $\sim1.2\times10^5$, or in $\sim1$ second with $\sim3.3\times10^8$~\cite{Litinski2019}.
Practical runtimes therefore call for massive \emph{parallel} distillation and a very large physical footprint.

\subsection{Resource Estimates for \SHA Preimage via Grover}
Amy et al.~\cite{Amy2017-SHA} deliver an end-to-end cost for SHA-2/3 preimage under the surface code, synthesizing reversible hash oracles and pricing them in surface-code time--area.
For \SHA they report depth $\approx 2^{153.8}$ code cycles and $\approx 2^{12.6}$ logical qubits, for a total volume $\approx 2^{166.4}$ logical-qubit-cycles~\cite{Amy2017-SHA}.
Grover's black-box bound for a 256-bit preimage reads $\Theta(2^{128})$ queries; the full fault-tolerant oracle inflates the bill by $\sim 2^{38}$ ($\sim 2.75\times10^{11}$) beyond that naïve count~\cite{Amy2017-SHA}.
Complementary perspectives include block-cipher resource estimates~\cite{Grassl2016GroverAES}, software-oriented oracle construction~\cite{Preston2022HashOracles}, and explicit time--space trade-offs for quantum search in symmetric cryptanalysis~\cite{Kim2018TimeSpace}.\footnote{The methodology extends to SHA-3 (FIPS~202~\cite{FIPS202}), though our oracles target SHA-256 exclusively.}

Gheorghiu and Mosca~\cite{Gheorghiu2019-Mosca} extend this approach across symmetric and hash-based schemes with surface-code assumptions that match current practice.
For Bitcoin's \dSHA~\cite{FIPS180-4,Nakamoto2008}, the overheads consume most of Grover's quadratic edge at realistic error rates and cycle times. Recovering that edge would require vast parallelism and high-throughput \emph{T}-state factories.

\subsection{Game-Theoretic Implications for Bitcoin Mining}
A quantum miner does not play the classical mining game.
Sattath~\cite{Sattath2020} shows that a Grover-enabled miner stops and measures as soon as a rival announces a block, which creates a race and lifts fork risk.
This strategic coupling, absent classically, threatens consensus.
Sattath outlines protocol-level fixes that desynchronize quantum miners.
Benkoczi et al.~\cite{BenkocziEtAl2022QuantumBitcoinMining} cast mining as a hybrid classical/quantum search problem and lay out a five-step workflow: classical precomputation, coherent nonce search, Grover amplification, measurement, and classical verification.
Nerem and Gaur~\cite{NeremGaur2023AdvantageousQuantumMining} then derive conditions under which a ``peaceful'' Grover miner beats a classical miner, emphasizing the cost per Grover step, the achievable query rate, and an optimal measurement horizon of roughly 16 minutes.
Bailey and Sattath~\cite{BaileySattath2024} identify a complementary attack vector: a quantum miner can manipulate block timestamps to inflate the epoch-wide difficulty adjustment, raising the cost for classical competitors while exploiting Grover's quadratic edge to absorb the increase more cheaply.
This difficulty-increase strategy shifts the threat from a single-block race to a multi-epoch economic squeeze.
Together, these papers establish the algorithmic and economic conditions for quantum mining, but they do not price a reversible double-\SHA oracle, fault-tolerant non-Clifford supply, or the fleet-scale qubit and power budgets required to realize those conditions.
We close that physical-cost gap.
Cojocaru et al.~\cite{Cojocaru2023PostQuantumBitcoin} analyze the post-quantum security of authenticated key-exchange protocols and note analogous deadline-constrained attack models, reinforcing that quantum threats to blockchain security extend beyond mining into the protocol layer.
Together, these results demonstrate that performance, incentives, and security intertwine tightly.

\subsection{Practical Limits of Quadratic Speedups}
Babbush et al.~\cite{Babbush2021} survey quadratic speedups in the fully error-corrected regime.
With modern surface-code costing they argue: a quadratic gain rarely crosses the practicality bar on early fault-tolerant hardware.
Cade et al.~\cite{Cade2023QuantifiedGrover} sharpen this conclusion with a non-asymptotic framework that captures exact constant prefactors in Grover's algorithm; their analysis confirms the crossover point where quantum search overtakes classical brute force lies far beyond near-term device scales, a finding our parametric sweep corroborates with explicit surface-code fleet numbers.
Brehm and Weggemans~\cite{BrehmWeggemans2026FTQCAdvantage} reach a parallel conclusion for lattice sieving: even under optimistic fault-tolerant assumptions, quadratic quantum speedups for the shortest-vector problem remain impractical, reinforcing the broader pattern that quadratic advantages rarely survive error-correction overhead.
For \SHA preimage, the constants dominate. The crucial ones are qubits, factories, and cycles. Classical hashing still wins on speed and on cost.

\subsection{Energy Estimation for Fault-Tolerant Quantum Computing}
Two recent works bracket the power-per-qubit uncertainty that dominates large-scale resource projections.
Parker and Vermeer~\cite{ParkerVermeer2023} adopt a policy-oriented, top-down approach: they factor energy per cryptanalytic key into the product of spacetime volume (qubit-days, drawn from Gidney--Eker{\aa} and earlier surface-code estimates) and average power per physical qubit.
Extrapolating from Google's Sycamore system and IBM's Osprey refrigerator, they arrive at $\sim6$~W/qubit for a future superconducting cryptanalytically relevant quantum computer (CRQC), yielding $\sim125$~MW total power and $\sim900$~MWh per RSA-2048 key. They frame those costs as nation-state-scale.

Fellous-Asiani et al.~\cite{FellousAsiani2023} pursue the opposite tack: a bottom-up ``Metric--Noise--Resource'' (MNR) framework that couples qubit noise models, concatenated error correction, multi-stage Carnot-efficient cryogenics, and control-electronics power into a single optimization.
Under aggressive but physically consistent assumptions, they derive $\sim1$--$2$~mW/qubit for large fault-tolerant workloads. That is three to four orders of magnitude below Parker's empirical extrapolation.
Applied to the same Gidney--Eker{\aa} 20-million-qubit surface-code Shor instance, their optimistic scenario projects only $\sim20$--$40$~kW total power and $\sim0.2$~MWh per key.
Relaxing cryogenic ideality or adding parasitic heat loads (Appendix~E of that work) pushes power back toward gigawatts, bracketing the conservative estimates.

For Bitcoin mining, the algorithm's intrinsic cost dwarfs this engineering uncertainty: even at 1~mW/qubit, a true-preimage fleet of $\sim10^{75}$ qubits would exceed any astrophysical power source.
We therefore adopt a conservative band ($3$--$12$~W/qubit depending on architecture) aligned with Parker's methodology, treating the Fellous-Asiani optimistic bound as a floor that does not qualitatively alter the Kardashev-scale conclusions.
More broadly, this body of literature exposes a gap between asymptotic query counts and physical feasibility: oracle design, non-Clifford throughput, and fault-tolerance overhead control the real cost.
We close that loop by specifying a mining search model, tallying reversible hash oracles, and mapping the resulting logical counts into surface-code fleets and energy budgets.

\subsection{Classical Mining Baselines: Hashrate and Power}
\label{sec:classical-baselines}

Comparing quantum fleet power against the incumbent network requires a compact classical cost model.
Bitcoin retargets the proof-of-work threshold so that blocks arrive every ten minutes on average.
Writing difficulty as \(D\), the canonical approximation for network hashrate reads~\cite{BitcoinWikiDifficulty2025}:
\begin{align}
H_{\mathrm{net}}(D)&\ \approx\ \frac{D\cdot 2^{32}}{600}\ \si{\hash\per\second},
\label{eq:btc-hashrate-from-difficulty}\\
H_{\mathrm{net}}^{(\mathrm{TH/s})}(D)&\ =\ \frac{D\cdot 2^{32}}{600\times 10^{12}}\ \si{\tera\hash\per\second}.
\notag
\end{align}
A fleet-average ASIC efficiency \(\eta\) in \(\si{\joule\per\tera\hash}\) converts hashrate to electrical power:
\begin{equation}
P_{\mathrm{net}}(D;\eta)\ =\ \eta\cdot H_{\mathrm{net}}^{(\mathrm{TH/s})}(D)\ \si{\watt}.
\label{eq:btc-power-from-difficulty}
\end{equation}
For~\(\eta\), we track a fleet-average trajectory from the Cambridge Bitcoin Electricity Consumption Index (CBECI)~\cite{CBECI2025,Cambridge2023ImprovedAssessment} and bracket it with device-level bands from representative ASICs: the Antminer S9~(\(\sim\!80\)~\si{\joule\per\tera\hash}), S19~(\(\sim\!29.5\)~\si{\joule\per\tera\hash}), and S21~(\(\sim\!17.5\)~\si{\joule\per\tera\hash})~\cite{BitmainS9SESpec2019,BitmainS19ProHydroSpecs2022,BitmainS21ImmersionSpecs2024,MicroBTM60Series2025}.
These baselines furnish the classical yardstick against which the quantum fleet power in \secref{sec:energy-kardashev} stands measured.

\section{Mining model and logical resource estimation}
\label{sec:logical-grover}

Every Grover iteration pays a non-Clifford tax set almost entirely by the reversible hash oracle. The comparator and diffusion together contribute less than $1\%$ of the $T$-count.
This section traces that tax from the Bitcoin protocol down to gate-level ledgers, producing the iteration budgets that drive the surface-code mapping in \secref{sec:physical-footprint}.
The oracle mirrors Bitcoin mining: a reversible \dSHA evaluation followed by a threshold comparison.
The \SHA primitives follow FIPS~180-4~\cite{FIPS180-4}; Appendix~\ref{app:rev-sha} specifies the reversible building blocks fed into the estimator.
The estimator tallies logical qubits, $T$ count, and $T$ depth; it then wraps the oracle into Grover's phase query and hands off the resulting costs to the surface-code compiler.

\begin{figure*}[t]
  \centering
  \resizebox{0.95\textwidth}{!}{% Grover mining pipeline schematic for Kardashev manuscript
% Improved for readability: larger fonts, cleaner layout, better spacing
\begin{tikzpicture}[
  font=\sffamily,
  % Main processing blocks - larger and more readable
  block/.style={
    draw=#1!60!black,
    line width=1.2pt,
    rounded corners=6pt,
    inner sep=12pt,
    align=center,
    text width=3.8cm,
    minimum height=2.4cm,
    fill=#1!15,
    font=\sffamily
  },
  % Thicker, more visible arrows
  arrow/.style={
    -{Latex[length=4mm, width=2.8mm]},
    line width=1.4pt,
    draw=gray!70
  },
  % Edge labels - larger and clearer
  edgelabel/.style={
    font=\sffamily\small,
    align=center,
    fill=white,
    inner sep=4pt,
    rounded corners=2pt
  },
  % Feedback arrows - more distinct
  feedbackarrow/.style={
    -{Latex[length=3.5mm]},
    line width=1.2pt,
    draw=gray!50,
    dashed,
    dash pattern=on 4pt off 2pt
  },
  feedbacklabel/.style={
    font=\sffamily\small\itshape,
    text=gray!70!black,
    fill=white,
    inner sep=3pt,
    rounded corners=2pt,
    align=center
  },
  % Stage labels above boxes
  stagelabel/.style={
    font=\sffamily\small,
    text=gray!60!black,
    above=3pt
  }
]

  % ===== TOP ROW: Main pipeline (left to right) =====
  \node[block=blue] (difficulty) {
    \textbf{\large Difficulty}\\[6pt]
    $D$, success rate\\[2pt]
    Marked-state density
  };

  \node[block=cyan, right=2.2cm of difficulty] (ledger) {
    \textbf{\large Hash Oracle}\\[6pt]
    Double-\SHA\\+ \RIPEMD\\[2pt]
    $\mathcal{O}_{\text{hash}}$
  };

  \node[block=green!70!black, right=2.2cm of ledger] (grover) {
    \textbf{\large Grover}\\[6pt]
    Iteration count $k$\\[2pt]
    $G^{k}\mathcal{O}_{\text{hash}}$
  };

  \node[block=orange, right=2.2cm of grover] (fleet) {
    \textbf{\large Fleet}\\[6pt]
    $N_{\text{fleet}}$ machines\\[2pt]
    Duty cycle
  };

  % ===== BOTTOM ROW =====
  \node[block=violet, below=2.0cm of $(ledger.south)!0.5!(grover.south)$] (surface) {
    \textbf{\large Surface Code}\\[6pt]
    Distance $d$, $T$-factories\\[2pt]
    Cycle time $\tau_{\text{cyc}}$
  };

  \node[block=red, below=2.0cm of fleet] (energy) {
    \textbf{\large Energy}\\[6pt]
    $P_{\text{fleet}}$ (TW)\\[2pt]
    Kardashev scale
  };

  % ===== MAIN FLOW ARROWS =====
  \draw[arrow] (difficulty.east) -- node[edgelabel, above] {$N{=}2^{256}$} (ledger.west);
  \draw[arrow] (ledger.east) -- node[edgelabel, above] {Gate counts} (grover.west);
  \draw[arrow] (grover.east) -- node[edgelabel, above] {$k$ iterations} (fleet.west);
  \draw[arrow] (fleet.south) -- node[edgelabel, right, pos=0.5] {Power} (energy.north);

  % Surface code feeds into fleet - clean diagonal with label
  \draw[arrow] (surface.east) -- node[edgelabel, above, pos=0.5] {Qubits, runtime} (fleet.south west);

  % ===== FEEDBACK LOOPS (dashed) =====
  % Grover -> Surface Code: straight down
  \draw[feedbackarrow] (grover.south) -- node[feedbacklabel, right=2pt] {Cycle budget} (surface.north);

  % Surface Code -> Difficulty: simple diagonal
  \draw[feedbackarrow] (surface.south west) -- node[feedbacklabel, below, pos=0.5] {Tuning} (difficulty.south east);

  % ===== STAGE LABELS (subtle, above each box) =====
  \node[stagelabel] at (difficulty.north) {Input};
  \node[stagelabel] at (ledger.north) {Oracle};
  \node[stagelabel] at (grover.north) {Algorithm};
  \node[stagelabel] at (fleet.north) {Scaling};
  \node[stagelabel, below=3pt] at (surface.south) {Error correction};
  \node[stagelabel, below=3pt] at (energy.south) {Output};

\end{tikzpicture}}
  \caption{End-to-end Grover mining estimator.  Difficulty and success targets feed the reversible hash oracle, Grover iteration planner, and surface-code lift; the resulting fleet sizing drives the power and energy tallies used in the Kardashev analysis.  Dashed arrows mark feedback between iteration budgets and surface-code parameters when runtime caps shrink.}
  \label{fig:grover-pipeline}
\end{figure*}
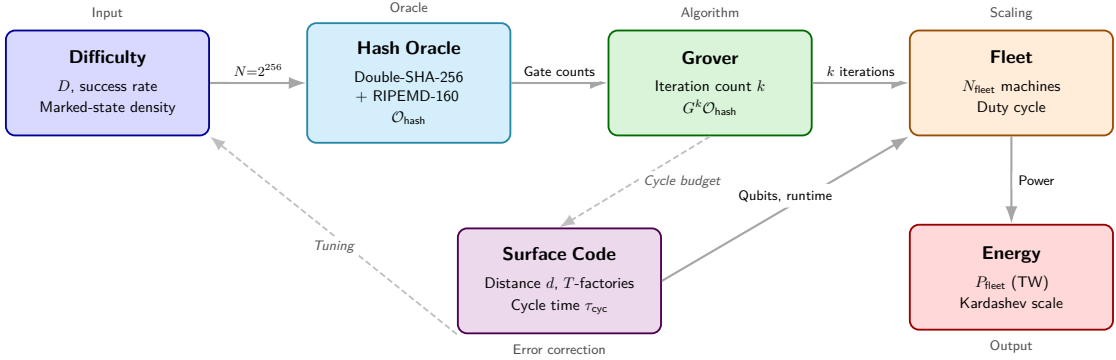

Figure~\ref{fig:grover-pipeline} summarizes the end-to-end pipeline from difficulty to power.

\subsection{Mining search space and oracle input model}
\label{sec:mining-input-model}

Bitcoin mining evaluates \(H(m)=\dSHA(m)\) on an 80-byte block header~\cite{Nakamoto2008}.
The header concatenates version, previous-block hash, Merkle root, timestamp, compact target (\texttt{nBits}), and a 32-bit nonce.
During a mining window a miner fixes the previous-block hash and the epoch-wide target and varies other fields to generate candidate headers.
The following definitions anchor the notation used throughout the paper.

\begin{center}
\fbox{%
\begin{minipage}{0.97\linewidth}
\textbf{Definitions: protocol parameters.}
\begin{description}
  \item[$T$ (target)] The proof-of-work predicate marks a header $m$ that satisfies $H(m) < T$.
  \item[$D$ (difficulty)] Bitcoin sets $T = T_1 / D$ with $T_1$ the difficulty-1 target.
  \item[$p$ (hit probability)] Model $H(m)$ as uniform over $\{0,1\}^{256}$, so $p=\Pr[H(m) < T]=T/2^{256}=(T_1/2^{256})/D\approx 2^{-32}/D$.
  \item[$b$ (difficulty bits)] Set $b := -\log_2 p$; a power-of-two target $T=2^{256-b}$ makes $b$ match the leading-zero bit count in $H(m)$.
\end{description}
\medskip
\textbf{Definitions: quantum search parameters.}
\begin{description}
  \item[$N=2^n$ (search size)] $n$ qubits label a domain of size $N=2^n$.
  \item[$M$ (marked inputs)] The marked count satisfies $M=pN$; when $n=256$ this gives $M=p2^{256}=T$.
  \item[$r$ (Grover iterations)] With $\theta=\arcsin\sqrt{M/N}$, use $r=\max\!\left\{1,\operatorname{round}\!\left(\frac{\pi}{4\theta}-\frac{1}{2}\right)\right\}\approx \pi/(4\sqrt{p})$ for $p\ll1$.
\end{description}
\end{minipage}}
\end{center}

Model the quantum search register as \(n\) qubits \(\lvert x\rangle\) that index the adjustable degrees of freedom supplied to the header oracle.
Two limiting cases bracket the choice of search register:
\begin{description}
  \item[Nonce-only (\(n=32\)).] Place the nonce field in superposition and treat the remaining 48 header bytes, including the Merkle root, as classical constants.
  \item[Nonce plus auxiliary freedom (\(n>32\)).] Extend the superposition with extra degrees of freedom such as an \texttt{extraNonce} inside the coinbase transaction, the timestamp, or version bits. Those variables propagate into the Merkle root and therefore into the header.
\end{description}

Unless stated otherwise, set \(n=256\) to model a miner that draws from a large header domain without exhausting degrees of freedom during a Grover run.
This work keeps the simplified oracle \(m\mapsto \dSHA(m)\) on the resulting header and does not include the reversible cost of constructing a consistent Merkle root from a transaction set.
This simplification matches the nonce-only model and any model that holds the Merkle root fixed during the quantum run.
For a block with \(N_{\mathrm{tx}}\) transactions, a protocol-faithful oracle would hash the coinbase transaction and then recompute the Merkle path to the root, adding roughly \(1+\lceil\log_2 N_{\mathrm{tx}}\rceil\) extra \dSHA evaluations per query.
Omitting this step understates oracle depth and width, so the mining footprints reported here remain optimistic whenever the superposition includes \texttt{extraNonce} or transaction selection.
To keep this sensitivity explicit, introduce a hash-work factor
\begin{equation}
  k_{\mathrm{hash}} := 1 + \alpha_{\mathrm{merkle}}\!\left(1+\left\lceil\log_2 N_{\mathrm{tx}}\right\rceil\right),
  \label{eq:khash-merkle}
\end{equation}
where \(\alpha_{\mathrm{merkle}}\in\{0,1\}\) toggles whether the coherent search changes the Merkle root during the run (\(\alpha_{\mathrm{merkle}}=0\) for the header-only baseline, \(\alpha_{\mathrm{merkle}}=1\) for protocol-faithful Merkle recomputation).
Nonce-only mining also admits midstate reuse for the first compression block.  Capture that engineering detail with \(\beta_{\mathrm{midstate}}\in\{1,1/2\}\), and set \(\beta_{\mathrm{midstate}}=1\) whenever \(\alpha_{\mathrm{merkle}}=1\).

Bitcoin's compact target encoding gives $T_1 = 2^{208} (2^{16}-1)$ (\texttt{0x00000000ffff0000\ldots}), so $T_1/2^{256} = 2^{-32}-2^{-48} \approx 2^{-32}$.
The estimator therefore evaluates
\(
M(n,D)=\max\!\left\{1,\min\!\left\{2^n,\frac{T_1}{D}\,2^{n-256}\right\}\right\},
\)
which preserves the $n=256$ mining baseline $M=T_1/D$ while keeping the nonce-only ($n=32$) and address-preimage ($n=160$) cases finite.
Grover's rotation angle is $\theta = \arcsin \sqrt{M/N}$, so the optimal number of iterations satisfies
\begin{equation}
\label{eq:grover-iterations}
r = \max\!\left\{1,\operatorname{round}\!\left(\frac{\pi}{4\theta} - \frac{1}{2}\right)\right\},
\end{equation}
with success probability $\sin^2((2r+1)\theta)$.
Here $\operatorname{round}(x)$ denotes round-to-nearest integer rounding (ties to even), matching the estimator implementation used in \secref{sec:physical-footprint}.
When $M$ drifts during a mining race, fixed-point amplitude amplification~\cite{Yoder2014FixedPoint} can replace this schedule at a constant-factor overhead in oracle calls.

\subsection{Ledger for the reversible \SHA oracle}

Using 32-bit Cuccaro--Draper--Kutin--Moulton (CDKM) ripple adders~\cite{Cuccaro2004} with Gidney's measurement-assisted
ANDs~\cite{Gidney2018Adder} and relative-phase Toffolis for boolean layers~\cite{Maslov2016RelativePhaseToffoli}, the logical
ledger per compression round reads
\begin{subequations}
\label{eq:round-ledger}
\begin{align}
N_{\mathrm{add}}(t) &=
\begin{cases}
7 & t < 16,\\
10 & t \ge 16,
\end{cases} \\
T_{\mathrm{bool}} &= 96, \\
\mathrm{CNOT}_{\Sigma}(t) &=
\begin{cases}
2 \cdot 96 & t < 16,\\
4 \cdot 96 & t \ge 16.
\end{cases}
\end{align}
\end{subequations}
Summing over the 64 rounds and the feed-forward stage gives the forward
\SHA compression ledger shown in Table~\ref{tab:sha-ledger}.  The Python
implementation exposes three adder choices: a baseline CDKM ripple, Gidney's
measurement-assisted scheduling for reduced depth, and a carry-save pipeline
that compresses the three additions in $T_1$ before a single ripple.  Our
automated verification suite asserts $1{,}800$
invocations of \(\mathrm{Add}_{32}\), $18{,}432$ boolean Toffolis, a $T$-count of
$3.04128\times 10^5$, and a footprint of $833$ logical qubits for the double
hash, while also checking the expected $T$-depth and $T$-count deltas for the
alternate adder models.  The ledger sits within $10\%$ of the reference
tally reported by Amy \emph{et al.}~\cite{Amy2017-SHA}. Our verification harness enforces that check at
runtime. The ledger also agrees with the order-of-magnitude
estimates in \eqnref{eq:rev-T-round}.  Standard Toffoli synthesis
adds three $T$ gates per Toffoli; the table records the resulting penalties in
parentheses.  An SHA-specific circuit optimization targeting $T$-depth can further compress the depth constant; see Lee~\emph{et~al.}~\cite{Lee2023SHA256TDepth}.  The estimator accepts such alternatives as drop-in oracle modules.

\begin{table}[t]
  \centering
  \caption{Logical ledger for one forward evaluation of \dSHA with configurable 32-bit adders.  ``Adders'' counts full-width modular additions; ``boolean'' counts relative-phase Toffolis from $\mathrm{Ch}$ and $\mathrm{Maj}$.  T-count deltas reflect the carry-save pipeline, while T-depth deltas report both Gidney scheduling (\(\Delta_G\)) and the carry-save pipeline (\(\Delta_{\mathrm{CS}}\)).  Parenthesized ``std'' terms give the additional T-count incurred when using standard Toffoli synthesis.}
  \label{tab:sha-ledger}
  \scriptsize
  \setlength{\tabcolsep}{3pt}
  \renewcommand{\arraystretch}{1.15}
  \resizebox{\linewidth}{!}{%
    \begin{tabular}{@{}lrrrrrr@{}}
      \toprule
      Stage & Adders & Boolean Toffolis & Total Toffolis & T-count & T-depth & CNOTs \\
      \midrule
      Compression (3\,$\times$) & \num{1776} & \num{18432} & \num{130320} & \makecell[r]{\num{301056}\\{\scriptsize$\Delta_{\mathrm{CS}}=-\num{24576}$}\\{\scriptsize$+\num{390960}\,\text{std}$}} & \makecell[r]{\num{112848}\\{\scriptsize$\Delta_G=-\num{53280}$}\\{\scriptsize$\Delta_{\mathrm{CS}}=-\num{65568}$}} & \num{398640} \\
      Feed-forward (3\,$\times$) & \num{24} & \num{0} & \num{1512} & \makecell[r]{\num{3072}\\{\scriptsize$+\num{4536}\,\text{std}$}} & \makecell[r]{\num{1512}\\{\scriptsize$\Delta_G=\Delta_{\mathrm{CS}}=-\num{720}$}} & \num{3768} \\
      \midrule
      \textbf{Double hash} & \num{1800} & \num{18432} & \num{131832} & \makecell[r]{\num{304128}\\{\scriptsize$\Delta_{\mathrm{CS}}=-\num{24576}$}\\{\scriptsize$+\num{395496}\,\text{std}$}} & \makecell[r]{\num{114360}\\{\scriptsize$\Delta_G=-\num{54000}$}\\{\scriptsize$\Delta_{\mathrm{CS}}=-\num{66288}$}} & \num{402408} \\
      \bottomrule
    \end{tabular}
  }
  \renewcommand{\arraystretch}{1.0}
\end{table}

\subsection{P2PKH hash pipeline}

When a user reuses a pay-to-public-key-hash (\PtwoPKH) address or an attacker recovers a public key from a spent output, the relevant oracle becomes \RIPEMD(\SHA(pubkey)) rather than the block-header double hash.
Bitcoin's \PtwoPKH addresses compose a single \SHA compression with \RIPEMD~\cite{Dobbertin1996RIPEMD160,BitcoinDevGuideTransactions}; Appendix~\ref{app:rev-sha} derives the reversible \RIPEMD circuit and Table~\ref{tab:ripemd-ledger} records the gate-level ledger.  Combining it with Table~\ref{tab:sha-ledger}, one forward evaluation of \RIPEMD(\SHA(pubkey)) consumes $600+650=1{,}250$ CDKM adders, $10{,}240$ boolean Toffolis, and $200{,}960$ $T$ gates.  Standard Toffoli synthesis adds another $266{,}970$ $T$ gates.  An additional $256$-qubit buffer holds the intermediate \SHA digest while the \RIPEMD branch executes, raising the peak width to $1{,}153$ logical qubits.

\begin{table}[t]
  \centering
  \caption{Comparison of reversible hash pipelines.  The \PtwoPKH oracle hashes a compressed public key through \SHA followed by \RIPEMD; the block-header oracle iterates \SHA twice.  The standard-Toffoli penalty counts the extra $T$ gates that standard Toffoli synthesis adds beyond relative-phase Toffolis.}
  \label{tab:p2pkh-ledger}
  \scriptsize
  \setlength{\tabcolsep}{3pt}
  \renewcommand{\arraystretch}{1.15}
  \resizebox{\linewidth}{!}{%
    \begin{tabular}{@{}lrrrrrrr@{}}
      \toprule
      Oracle & Logical qubits & Adders & Boolean Toffolis & Total Toffolis & T-count & Std.\ Toffoli penalty & CNOTs \\
      \midrule
      $\RIPEMD(\SHA(\mathrm{pubkey}))$ & \num{1153} & \num{1250} & \num{10240} & \num{88990} & \num{200960} & \num{266970} & \num{244378} \\
      $\dSHA(\mathrm{header})$ & \num{833} & \num{1800} & \num{18432} & \num{131832} & \num{304128} & \num{395496} & \num{402408} \\
      \bottomrule
    \end{tabular}
  }
  \renewcommand{\arraystretch}{1.0}
\end{table}

The \PtwoPKH oracle trades a slight reduction in $T$ count for a broader footprint: the \RIPEMD stage dominates the qubit tally, while the double-\SHA block-header oracle spends more non-Clifford gates in its boolean layers.

\subsection{Grover iteration cost}

One Grover iteration with a clean oracle entails four steps: compute the double hash, apply a $256$-bit less-than test, flip the sign of the auxiliary qubit, and uncompute all work registers.
In $T$-count terms the iteration yields
\begin{equation}
\Toracle = 2\,k_{\mathrm{hash}}\,\beta_{\mathrm{midstate}}\times 304{,}128 + 2 \times 1{,}024 + \Tdiff,
\label{eq:tcount-grover}
\end{equation}
where the leading term comes from the reversible hash work (Table~\ref{tab:sha-ledger}), the second term from a CDKM-style constant adder that implements the comparison, and the final term from the diffusion reflection.
A relative-phase Toffoli ladder synthesizes the diffusion. Its non-Clifford cost obeys $\Tdiff \approx 8(n-2)$ for an $n$-qubit search register with one clean ancilla. That load stays small but non-zero relative to the oracle itself.
Gate counts, $T$ depth, and qubit width all live in the estimator's programmatic interface, enabling rapid what-if studies.

The comparator follows a lexicographic subtraction of the difficulty target.
Each 32-bit slice of the $256$-qubit search register feeds a CDKM
ripple adder that accumulates the borrow into a single flag marking whether the
input fell below the classical threshold.  The workspace remains dirty; replaying
the sequence uncomputes it and produces the $2 \times 1{,}024$
contribution in \eqnref{eq:tcount-grover}.  The estimator also offers Gidney's measurement-assisted variant of
the CDKM ripple~\cite{Gidney2018Adder}: it preserves
the same Toffoli and $T$ counts but cuts the $T$-depth of each chunk from $63$
layers to roughly $33$, matching the ``$word\_size + 1$'' scheduling of
Gidney's trick.

Given an iteration budget $r$, the total non-Clifford demand reads
\begin{equation}
\Ttot = r \times \Toracle,
\label{eq:tcount-total}
\end{equation}
and likewise for $T$-depth and Clifford counts.  These aggregates feed the
surface-code mapping in \secref{sec:physical-footprint}.

\begin{table}[t]
  \centering
  \caption{Sensitivity of the mining estimates to protocol fidelity and
    engineering knobs.  For $n=256$ the comparator and diffusion terms in
    \eqnref{eq:tcount-grover} contribute $<1\%$ of \(\Toracle\), so the
    multiplier \(k_{\mathrm{hash}}\beta_{\mathrm{midstate}}\) tracks the full
    \(T\)-count and \(T\)-depth to percent accuracy.  Converting physical qubits
    into wall-plug power multiplies by a per-qubit budget \(P_q\); our baseline
    architecture anchors span \(P_q\in[10^{-3},12]\,\mathrm{W/qubit}\) and therefore add
    up to a \(10^{4}\) uncertainty.}
  \label{tab:mining-knob-sensitivity}
  \setlength{\tabcolsep}{4pt}
  \renewcommand{\arraystretch}{1.15}
  \begin{tabular}{lcccc}
    \toprule
    Setting & \(\alpha_{\mathrm{merkle}}\) & \(N_{\mathrm{tx}}\) & \(\beta_{\mathrm{midstate}}\) & \(\times\) on \(k_{\mathrm{hash}}\beta_{\mathrm{midstate}}\) \\
    \midrule
    M0 header-only baseline & 0 & --- & 1 & 1 \\
    M0 + midstate reuse & 0 & --- & $1/2$ & $1/2$ \\
    M1 Merkle recomputation & 1 & 512 & 1 & 11 \\
    M1 Merkle recomputation & 1 & 2048 & 1 & 13 \\
    M1 Merkle recomputation & 1 & 8192 & 1 & 15 \\
    \midrule
    Power-per-qubit budget & \multicolumn{3}{c}{\(P_q\in[10^{-3},12]\,\mathrm{W/qubit}\)} & $10^{4}$ \\
    \bottomrule
  \end{tabular}
  \renewcommand{\arraystretch}{1.0}
\end{table}

\section{The physical footprint of quantum Bitcoin mining}
\label{sec:physical-footprint}
Fault tolerance drives the footprint.  Logical counts only scratch the surface:
a viable preimage attack must run on a fault-tolerant substrate.  Specializing
to the surface code, we use the standard phenomenological fit
\begin{equation}
\label{eq:surface-logical}
p_L \approx 0.1\, (100 p_{\mathrm{phys}})^{(d+1)/2},
\end{equation}
which captures the exponential suppression of logical faults with distance
below the $\sim1\%$ threshold~\cite{Fowler2012SurfaceCodes,Litinski2019}.  Here
$p_{\mathrm{phys}}$ denotes the two-qubit error rate and $d$ the code distance.
Many surface-code analyses report the scaling
$p_L\sim A (p_{\mathrm{phys}}/p_{\mathrm{th}})^{(d+1)/2}$; choosing
$p_{\mathrm{th}}\approx 1\%$ and $A\approx 0.1$ yields the numerical form in
\eqnref{eq:surface-logical}.  Decoder choices and noise biases shift $A$ and
$p_{\mathrm{th}}$ by order-one factors, so we treat \eqnref{eq:surface-logical}
as an order-of-magnitude fit.

To pick $d$, allocate a run-level logical-failure budget $p_{\mathrm{run}}=0.01$
and distribute it uniformly across the $\Ttot$ non-Clifford locations.  This
choice keeps logical faults below the success thresholds we target (up to
$P_t=0.99$), and a decade shift in $p_{\mathrm{run}}$ moves $d$ by only
$O(2)$ under \eqnref{eq:surface-logical}.
Treating each such location as an independent fault opportunity gives the
union-bound proxy $p_{\mathrm{run}}\lesssim \Ttot p_L$, so we target
\begin{equation}
\label{eq:logical-budget-tcount}
p_L \le p_{\mathrm{fail}}:=\frac{0.01}{\Ttot}.
\end{equation}
This $T$-count normalization follows the standard resource-estimation picture
in which $T$-state distillation and injection dominate the spacetime volume and
therefore dominate the logical-failure budget~\cite{GidneyFowler2019Factories,Litinski2019}.
For each logical qubit we assume a square patch of $2 d^2$ physical qubits and
a code cycle of $1\,\mu\mathrm{s}$, matching the scenarios in Litinski's
surface-code atlas~\cite{Litinski2019}.

Logical patches alone account for a small fraction of the physical machine;
the bulk of the silicon goes to magic-state distillation.
$T$-state distillation dominates the non-Clifford supply.  Following the
15-to-1 factory construction, a single factory occupies
$1.25 d^2$ physical qubits and emits one high-fidelity $T$ state every
$10 d$ code cycles.  Denote by $t_{\mathrm{logical}}$ the logical runtime,
the product of the Grover program's $T$-depth and the code cycle time
$\tau$.  The estimator sizes the factory farm to meet the average throughput
demand
\begin{equation}
\label{eq:factory-demand}
R_T = \frac{\Ttot}{t_{\mathrm{logical}}},
\end{equation}
The number of factories is then
$N_{\mathrm{fac}} = \lceil R_T / (1 / (10 d \tau)) \rceil$, where $\tau$ is the
cycle time.  Physical qubit totals report the sum of data-patch qubits and
factory qubits; runtimes are the logical depth multiplied by $\tau$.

Table~\ref{tab:failure-budget-sensitivity} tests a more conservative failure
budget that scales with logical spacetime volume rather than $T$ count.  Let
$Q_{\mathrm{tot}}$ denote the program's logical-qubit footprint and define the
volume proxy $V=Q_{\mathrm{tot}}(t_{\mathrm{logical}}/\tau)$, the number of
distance-$d$ patch-cycles executed across the run.  Replacing $\Ttot$ by $V$ in
\eqnref{eq:logical-budget-tcount} targets $p_L \le 0.01/V$.  This change raises
code distance by a few units and inflates per-machine footprints (and therefore
fleet footprints) by less than a factor of two across the architectures and
runtime caps that drive Figures~\ref{fig:fleet-heatmap} and~\ref{fig:fleet-tradeoff}.
\begin{table}[t]
  \centering
  \caption{Sensitivity of fault-tolerant mining footprints to the failure-budget
    proxy.  The baseline uses the $T$-count budget \eqnref{eq:logical-budget-tcount};
    the volume proxy replaces $\Ttot$ by $V=Q_{\mathrm{tot}}(t_{\mathrm{logical}}/\tau)$
    and targets $p_L \le 0.01/V$.  Values report one-machine totals for the mining
    oracle with $Q_{\mathrm{tot}}=833$ logical qubits.}
  \label{tab:failure-budget-sensitivity}
  \begin{tabular}{llccc}
    \toprule
    Architecture & $t_{\mathrm{cap}}$ (s) & $d$ (T$\rightarrow$V) & $Q_{\mathrm{machine}}$ (T$\rightarrow$V) & $\times$ \\
    \midrule
    Superconducting SC & $60$ & $19\rightarrow23$ & \num{8.29e5}$\rightarrow$\num{1.29e6} & \num{1.55} \\
    Neutral Atom SC & $60$ & $13\rightarrow17$ & \num{4.25e5}$\rightarrow$\num{7.65e5} & \num{1.80} \\
    Ion Trap SC & $60$ & $9\rightarrow11$ & \num{2.27e5}$\rightarrow$\num{3.46e5} & \num{1.53} \\
    Superconducting SC & $600$ & $21\rightarrow25$ & \num{1.04e6}$\rightarrow$\num{1.56e6} & \num{1.50} \\
    Neutral Atom SC & $600$ & $15\rightarrow19$ & \num{5.81e5}$\rightarrow$\num{9.80e5} & \num{1.69} \\
    Ion Trap SC & $600$ & $9\rightarrow11$ & \num{2.27e5}$\rightarrow$\num{3.46e5} & \num{1.53} \\
    \bottomrule
  \end{tabular}
\end{table}

Table~\ref{tab:example-footprint} lists the output of the estimator for a
difficulty-1 mining search with $n=256$ and the same physical parameters.
Sec.~\ref{sec:mining-input-model} specifies the mining input model and the
simplified block-header oracle behind this choice.
Because $M_1 \approx 2^{224}$ marked states exist at difficulty 1, the optimal
Grover budget is modest ($\sim5.1\times10^4$ iterations) and the surface-code
footprint lands in the million-qubit regime.  Even this ``easy'' instance
requires hundreds of simultaneous $T$ factories, highlighting the gulf between
logical resource counts and physical deployments.  Accounting for
$T_{\mathrm{diff}}(256) = 2.0\times 10^3$ nudges the factory tally in
Table~\ref{tab:example-footprint} to $6.1\times 10^2$, illustrating that the
diffusion reflection contributes a small but non-negligible distillation load.
Harder settings, such as a
single-solution preimage, scale back to the astronomical runtimes reported by
Amy \emph{et al.} and the factory-heavy layouts in Litinski's atlas.

\subsection{P2PKH preimage footprint}

Switching to the \PtwoPKH oracle trades the 256-bit comparator for a 160-bit
threshold while inheriting the wider logical core of the RIPEMD stage.  The
estimator therefore tracks $1{,}346$ logical qubits per iteration once the
search register and comparison ancillas join the $1{,}153$-qubit hash pipeline.
Table~\ref{tab:p2pkh-footprint} records the surface-code footprint for a
single-solution target ($D = M_1$) at $p_{\mathrm{phys}} = 10^{-3}$ and
$n=160$.  Grover's schedule reaches $9.5\times10^{23}$ iterations, inflating
the code distance to $d=61$ and pushing the logical runtime to
$1.5\times10^{23}$ seconds, which is roughly $5\times10^{15}$ years.  These numbers
echo the mining oracle's conclusion: even with the narrower comparator, the
preimage attack demands planetary-scale hardware.

\begin{table}[t]
  \centering
  \caption{Surface-code footprint for a single-solution \PtwoPKH preimage
  attack with $n=160$, $p_{\mathrm{phys}}=10^{-3}$, $\tau=1\,\mu\mathrm{s}$.}
  \label{tab:p2pkh-footprint}
  \begin{tabular}{lrr}
    \toprule
    Quantity & Value & Source \\
    \midrule
    Logical qubits $Q_{\mathrm{tot}}$ & $1{,}346$ & Oracle + ancillas \\
    Grover iterations $r$ & $9.5\times 10^{23}$ & \eqnref{eq:grover-iterations} \\
    Code distance $d$ & $61$ & \eqnref{eq:surface-logical} \\
    Data qubits & $1.0\times 10^{7}$ & $2 d^2 Q_{\mathrm{tot}}$ \\
    $T$ factories & $1.5\times 10^{3}$ & \eqnref{eq:factory-demand} \\
    Factory qubits & $7.2\times 10^{6}$ & $1.25 d^2 N_{\mathrm{fac}}$ \\
    Total qubits & $1.7\times 10^{7}$ & Sum \\
    Runtime & $1.5\times 10^{23}\,\mathrm{s}$ & $t_{\mathrm{logical}}$ \\
    \bottomrule
  \end{tabular}
\end{table}

\begin{table}[t]
  \centering
  \caption{Example surface-code footprint for difficulty-1 Bitcoin mining with
  $n=256$, $p_{\mathrm{phys}}=10^{-3}$, $\tau=1\,\mu\mathrm{s}$.}
  \label{tab:example-footprint}
  \begin{tabular}{lrr}
    \toprule
    Quantity & Value & Source \\
    \midrule
    Grover iterations $r$ & $5.1\times 10^4$ & \eqnref{eq:grover-iterations} \\
    Code distance $d$ & $23$ & \eqnref{eq:surface-logical} \\
    Data qubits & $1.2\times 10^6$ & $2 d^2 Q_{\mathrm{tot}}$ \\
    $T$ factories & $614$ & \eqnref{eq:factory-demand} \\
    Factory qubits & $4.1\times 10^5$ & $1.25 d^2 N_{\mathrm{fac}}$ \\
    Total qubits & $1.6\times 10^6$ & Sum \\
    Runtime & $8.0\times 10^3\,\mathrm{s}$ & $t_{\mathrm{logical}}$ \\
    \bottomrule
  \end{tabular}
\end{table}

Runtime caps and target success probabilities set fleet scale.
Propagation delays in Nakamoto consensus impose an operational race window; we encode that deadline with a per-machine wall-clock cap $t_{\mathrm{cap}}$.
Given a per-machine wall-clock budget $t_{\mathrm{cap}}$, define the
time-capped Grover iteration count by
\begin{equation}
\label{eq:grover-rcap}
r_{\mathrm{cap}}=\min\!\left\{r,\left\lfloor\frac{t_{\mathrm{cap}}}{t_{\mathrm{iter}}}\right\rfloor\right\},
\end{equation}
where $r$ comes from \eqnref{eq:grover-iterations} and $t_{\mathrm{iter}}$
denotes the wall-clock time of one Grover iteration on a single machine,
obtained by
multiplying the iteration's $T$-depth by the code cycle time $\tau$.
The single-machine success probability is then
\begin{equation}
\label{eq:single-machine-success}
P_1=\sin^2\!\bigl((2r_{\mathrm{cap}}+1)\theta\bigr).
\end{equation}
Assuming independent machines, a fleet of $N_{\mathrm{machines}}$ miners
achieves success probability
\begin{equation}
\label{eq:fleet-success}
P_t=1-(1-P_1)^{N_{\mathrm{machines}}},
\end{equation}
so
\begin{equation}
\label{eq:fleet-machines}
N_{\mathrm{machines}}=\left\lceil\frac{\ln(1-P_t)}{\ln(1-P_1)}\right\rceil,
\end{equation}
and the fleet qubit demand follows $Q_{\mathrm{fleet}}=N_{\mathrm{machines}}Q_{\mathrm{machine}}$.

\begin{figure}[t]
  \centering
  \IfFileExists{figures/fleet_heatmap.pdf}{%
    \includegraphics[width=0.95\columnwidth]{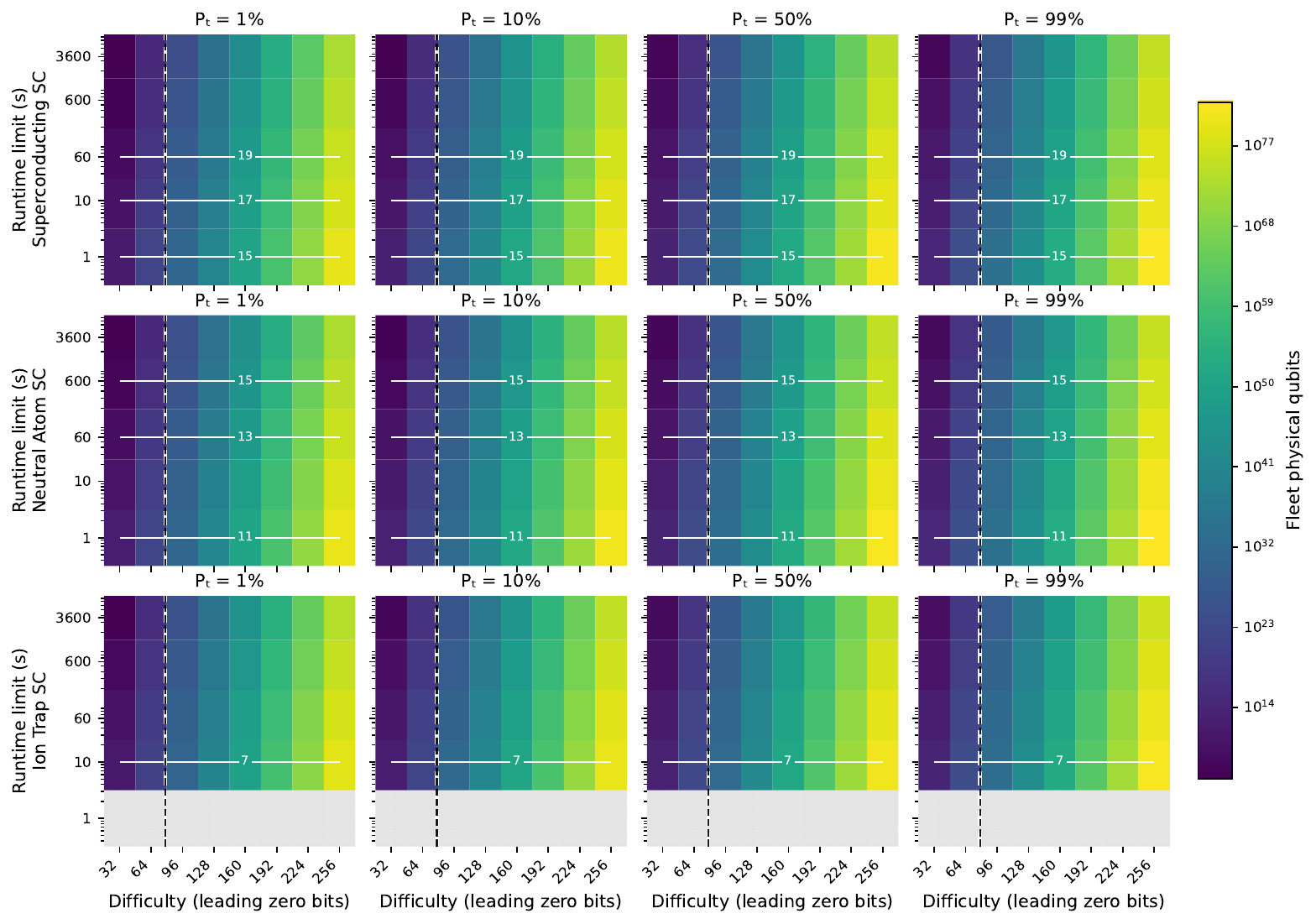}%
  }{%
    \fbox{\parbox{0.9\columnwidth}{\centering
      Figure~\ref{fig:fleet-heatmap} visualizes the fleet qubit
      requirements described in \secref{sec:physical-footprint}.}}%
  }
  \caption{Fleet qubits required to mine at various difficulty, runtime, and
    success targets for three surface-code architectures.  To read a cell,
    choose difficulty bits on the horizontal axis and a runtime cap on the
    vertical axis, then read the color as the fleet physical-qubit demand; grey
    cells mark infeasible settings where a single Grover iteration already
    exceeds the wall-clock cap, and white contours label code distance.  Each
    cell evaluates $r_{\mathrm{cap}}$ via \eqnref{eq:grover-rcap} and sizes
    fleets via \eqnref{eq:fleet-success}--\eqnref{eq:fleet-machines} with $P_1$
    from \eqnref{eq:single-machine-success}. Rows vary the hardware platform;
    columns fix the target success probability~$P_t$.  A dashed vertical marker
    highlights the Bitcoin mainnet difficulty on 2025-01-01 ($D\approx
    1.1\times10^{14}$, $b\approx78.6$)~\cite{BlockchainComDifficulty,BlockchainComChartsAPI}.}
  \label{fig:fleet-heatmap}
\end{figure}

\begin{figure}[t]
  \centering
  \IfFileExists{figures/fleet_tradeoff_curves.pdf}{%
    \includegraphics[width=0.95\columnwidth]{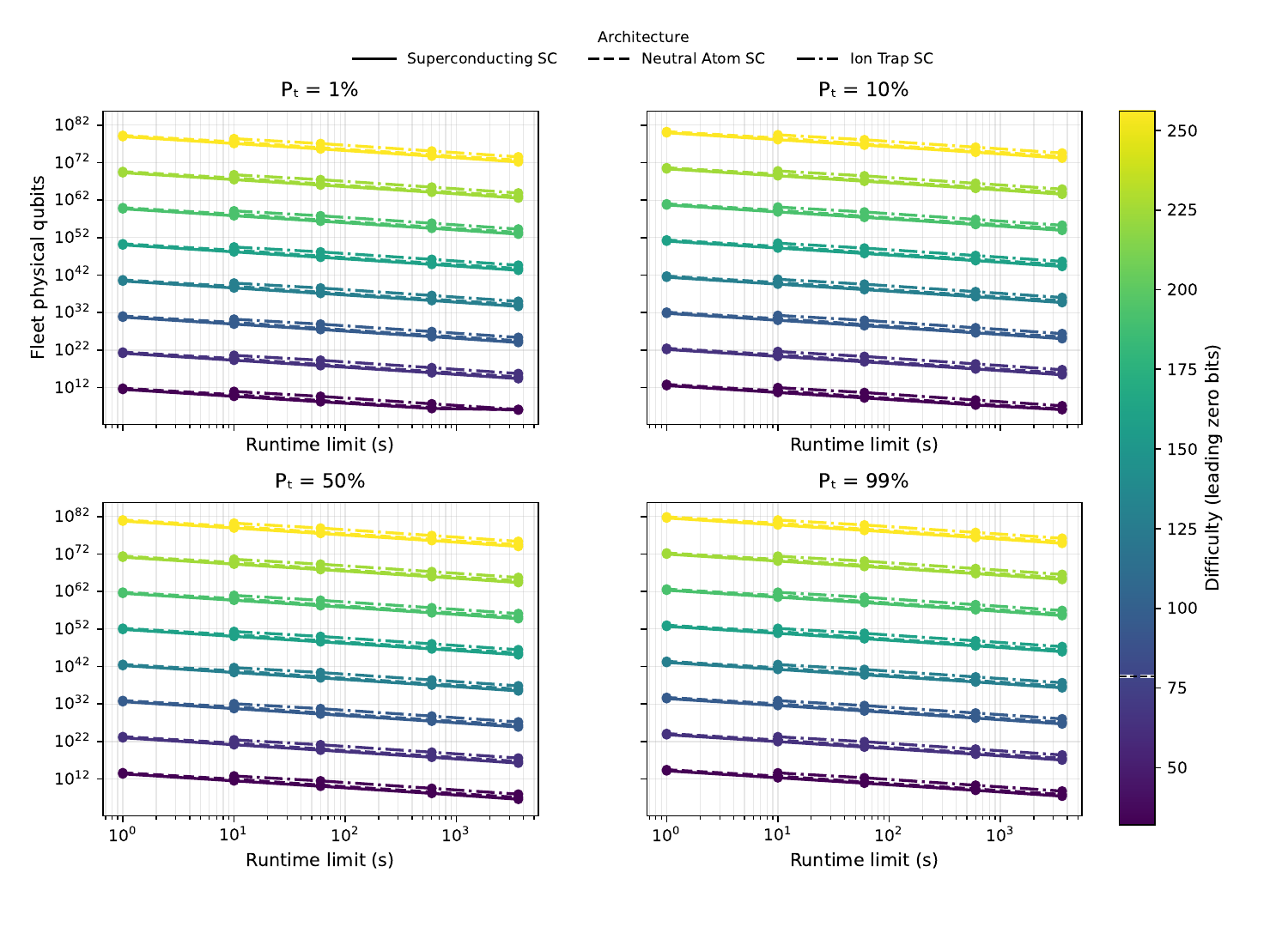}%
  }{%
    \fbox{\parbox{0.9\columnwidth}{\centering
      Figure~\ref{fig:fleet-tradeoff} reports runtime--qubit
      trade-offs drawn from the estimator sweep.}}%
  }
  \caption{Runtime--qubit Pareto fronts extracted from the same sweep as Figure~\ref{fig:fleet-heatmap}.  To read a point, pick a runtime cap on the horizontal axis and follow the curve for a chosen difficulty (color) and architecture (line style) to the fleet-qubit demand on the vertical axis.  Each panel fixes the target success probability~$P_t$ and evaluates \eqnref{eq:grover-rcap} and \eqnref{eq:fleet-success}--\eqnref{eq:fleet-machines}; missing segments correspond to infeasible settings where a single Grover iteration exceeds the runtime cap.  A dashed marker on the difficulty colorbar highlights the Bitcoin mainnet difficulty on 2025-01-01 ($b\approx78.6$)~\cite{BlockchainComDifficulty,BlockchainComChartsAPI}.}
  \label{fig:fleet-tradeoff}
\end{figure}

To chart how these footprints explode away from the difficulty-1 toy instance,
we sweep across runtime budgets, success thresholds, and surface-code
architectures using the estimator introduced above.  For every configuration we
solve \eqnref{eq:grover-iterations} and \eqnref{eq:grover-rcap} to fix the
Grover depth, fold the resulting $T$-count through \eqnref{eq:tcount-total},
and then size code distances and factory farms via Eqs.~\eqref{eq:surface-logical}
and~\eqref{eq:factory-demand}.  Finally, \eqnref{eq:single-machine-success}
and \eqnref{eq:fleet-machines} set the fleet size for a target $P_t$.  The
resulting data populate Figures~\ref{fig:fleet-heatmap} and~\ref{fig:fleet-tradeoff}.
Figure~\ref{fig:fleet-heatmap} highlights how architecture choices and performance targets inflate the required fleet size.
Grey cells flag runtime caps where even a single Grover iteration already exceeds the wall-clock budget, rendering those configurations infeasible.
White contour lines track code distance~$d$; higher distances pair with larger magic-state factories and correspondingly higher power draw.
Comparing columns shows how raising the success probability from $P_t=50\%$ to $P_t=99\%$ multiplies fleet size by roughly an order of magnitude across every architecture.

Slicing those evaluations by runtime and qubit budgets yields the Pareto view in Figure~\ref{fig:fleet-tradeoff}. The log-scale heatmaps color the total physical qubits across the required machine fleet, while contours annotate the underlying code distance. Greyed cells indicate that a single machine cannot execute even one Grover iteration within the runtime cap, forcing effectively infinite hardware.

\begin{table}[t]
  \centering
  \caption{Surface-code architecture assumptions for Figures~\ref{fig:fleet-heatmap}--\ref{fig:fleet-tradeoff}. Each triplet lists the code cycle time $\tau$, layout factor $\lambda$ (the spacelike expansion applied to data patches), and per-gate physical error rate $p_{\mathrm{phys}}$ used in the sweep.}
  \label{tab:sc-architectures}
  \begin{tabular}{lccc}
    \toprule
    Architecture & $\tau$ (code cycle) & $\lambda$ (layout) & $p_{\mathrm{phys}}$ \\
    \midrule
    Superconducting SC & $1\,\mu\mathrm{s}$ & $2.0$ & $1.0\times 10^{-3}$ \\
    Neutral Atom SC & $2\,\mu\mathrm{s}$ & $2.5$ & $5.0\times 10^{-4}$ \\
    Ion Trap SC & $10\,\mu\mathrm{s}$ & $3.0$ & $1.0\times 10^{-4}$ \\
    \bottomrule
  \end{tabular}
\end{table}

Table~\ref{tab:scenario-scaling} distils those sweeps into representative runtime--success combinations for the superconducting surface-code point. With a ten-minute wall-clock cap, the single-solution regime remains firmly in the $10^{75.6}$ fleet-qubit range for a 50\% hit rate and grows by nearly a decade when $P_t=99\%$. Partial preimages shrink the hardware bill dramatically yet still demand Kardashev-scale machines: a $2^{32}$-marked state rare preimage needs about $10^{66}$ qubits, moderately dense bins sit near $10^{46.7}$, and even $2^{224}$ marked states consume roughly $10^{8.2}$ qubits once distillation farms are counted. Tightening the runtime budget to 60 seconds shifts each row up by roughly two decades in machine count (e.g., the true-preimage line jumps from $10^{69.5}$ to $10^{71.5}$ machines, the densest partial case climbs from $10^{2.0}$ to $10^{4.0}$), underscoring how runtime caps and marked-state counts jointly shape fleet sizing.

\begin{table}[t]
  \centering
  \caption{Representative fleet scaling at a ten-minute runtime cap ($600\,$s) for the superconducting surface-code architecture. Values report the $\log_{10}$ of the required machine count ($N_{\mathrm{machines}}$, from \eqnref{eq:fleet-success}--\eqnref{eq:fleet-machines} with $P_1$ from \eqnref{eq:single-machine-success} and $r_{\mathrm{cap}}$ from \eqnref{eq:grover-rcap}) and total fleet qubits ($Q_{\mathrm{fleet}}$) for the most hardware-efficient configuration in each marked-state regime.}
  \label{tab:scenario-scaling}
  \begin{tabular}{lcccc}
    \toprule
    Regime & $\log_2\lvert S\rvert$ & $P_t$ & $\log_{10} N_{\mathrm{machines}}$ & $\log_{10} Q_{\mathrm{fleet}}$ \\
    \midrule
    True preimage (1 marked state) & $0$ & $0.50$ & $69.47$ & $75.58$ \\
    True preimage (1 marked state) & $0$ & $0.99$ & $70.29$ & $76.40$ \\
    Partial ($\le 2^{32}$ states) & $32$ & $0.50$ & $59.84$ & $65.95$ \\
    Partial ($\le 2^{32}$ states) & $32$ & $0.99$ & $60.66$ & $66.77$ \\
    Partial ($2^{33}$--$2^{96}$ states) & $96$ & $0.50$ & $40.57$ & $46.68$ \\
    Partial ($2^{33}$--$2^{96}$ states) & $96$ & $0.99$ & $41.39$ & $47.51$ \\
    Partial ($> 2^{96}$ states) & $224$ & $0.50$ & $2.04$ & $8.15$ \\
    Partial ($> 2^{96}$ states) & $224$ & $0.99$ & $2.86$ & $8.97$ \\
    \bottomrule
  \end{tabular}
\end{table}

Figure~\ref{fig:fleet-tradeoff} emphasizes the steep runtime penalties for constraining the hardware footprint.
Sliding from a 600-second to a 60-second deadline shifts every Pareto point upward by more than a decade of physical qubits, illustrating the severe cost of demanding faster Grover searches.
Marker styles distinguish architectures: superconducting surface-code machines anchor the fastest curves, neutral-atom arrays occupy an intermediate band, and ion-trap fleets trail by roughly an order of magnitude in runtime at matched qubit budgets.

Two cross-checks anchor these outputs.
First, the forward $T$-count produced by the estimator sits within $10\%$ of
the Amy~\emph{et al.}\ reference for the same SHA-256 oracle, confirming that
the reversible decomposition introduces no unexpected overhead.
Second, the Grover iteration formula reproduces the textbook
amplitude-amplification success probability to machine precision across the
full range of marked-state fractions.
The estimator emits both human-readable tables and machine-parseable JSON/CSV,
feeding directly into the Kardashev-scale energy analysis of
\secref{sec:energy-kardashev}.

\section{Energy and Kardashev scale}
\label{sec:energy-kardashev}

Fleet qubits translate directly into an energy bill.
This section converts the fleet sizes from \secref{sec:physical-footprint} into wall-plug power and compares that demand against the classical mining baselines from \secref{sec:classical-baselines} and Kardashev's Type~I/II/III thresholds.~\cite{Kardashev1964}

\begin{figure*}[t]
  \centering
  \IfFileExists{figures/hashrate_history.pdf}{%
    \includegraphics[width=0.92\textwidth]{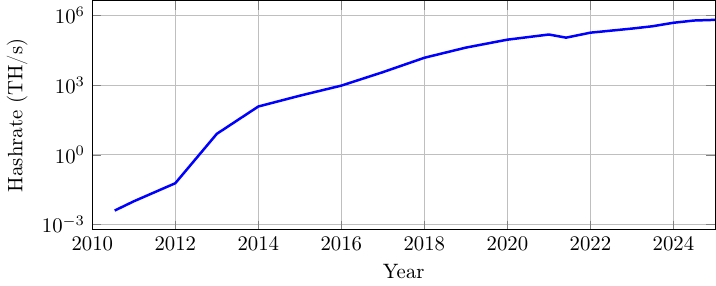}%
  }{%
    \IfFileExists{figures/hashrate_history.tikz}{%
      % !TEX root = ../kardashev.tex
\begin{tikzpicture}
  \begin{axis}[
    width=\linewidth, height=0.45\linewidth,
    xlabel={Year}, ylabel={Hashrate (TH/s)},
    xmin=2010, xmax=2025, xtick={2010,2012,2014,2016,2018,2020,2022,2024},
    xticklabel style={/pgf/number format/fixed, /pgf/number format/1000 sep={}},
    ymode=log,
    grid=both,
  ]
    \addplot+[mark=none, very thick] table [x=year_decimal, y=th_s, col sep=comma] {data/btc_hashrate_decimal.csv};
  \end{axis}
\end{tikzpicture}%
    }{%
      \fbox{\parbox{0.9\linewidth}{\centering Hashrate history (CSV missing).}}%
    }%
  }%
  \caption{Full-history Bitcoin network hashrate from Blockchain.com (7-day smoothing as provided).}
  \label{fig:btc-hashrate-history}
\end{figure*}

\begin{figure*}[t]
  \centering
  \IfFileExists{figures/power_vs_difficulty.pdf}{%
    \includegraphics[width=0.92\textwidth]{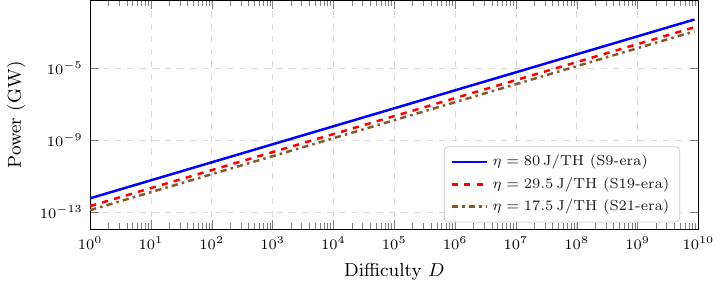}%
  }{%
    \IfFileExists{figures/power_vs_difficulty.tikz}{%
      \begin{tikzpicture}
  \begin{axis}[
    width=0.98\linewidth, height=0.45\linewidth,
    xlabel={Difficulty $D$}, ylabel={Power (\si{\giga\watt})},
    xmode=log,
    ymode=log,
    xmin=1, xmax=1e10,
    enlarge x limits=false,
    grid=major,
    major grid style={dashed, gray!30},
    legend pos=south east,
    legend cell align=left,
    legend style={font=\scriptsize, fill=white, fill opacity=0.9, draw=gray!40, rounded corners=2pt},
    tick label style={font=\scriptsize},
    label style={font=\small},
    axis background/.style={fill=white},
  ]
    % CSV columns: D, eta_JperTH, P_W
    \addplot+[mark=none, very thick, color=blue] table [x=D, y expr=\thisrow{P_W}/1e9, col sep=comma] {data/power_eta_80.csv};
    \addlegendentry{$\eta=80\,\si{J/TH}$ (S9-era)}
    \addplot+[mark=none, very thick, color=red, dashed] table [x=D, y expr=\thisrow{P_W}/1e9, col sep=comma] {data/power_eta_29_5.csv};
    \addlegendentry{$\eta=29.5\,\si{J/TH}$ (S19-era)}
    \addplot+[mark=none, very thick, color=brown!70!black, dashdotted] table [x=D, y expr=\thisrow{P_W}/1e9, col sep=comma] {data/power_eta_17_5.csv};
    \addlegendentry{$\eta=17.5\,\si{J/TH}$ (S21-era)}
  \end{axis}
\end{tikzpicture}%
    }{%
      \fbox{\parbox{0.9\linewidth}{\centering Power vs.\ difficulty (CSVs missing).}}%
    }%
  }%
  \caption{Network power as a function of difficulty under three efficiency tracks.}
  \label{fig:btc-power-vs-difficulty}
\end{figure*}

Figures~\ref{fig:btc-hashrate-history} and~\ref{fig:btc-power-vs-difficulty} summarize the classical network: full-history hashrate and network power as a function of difficulty under the ASIC efficiency tracks defined in \secref{sec:classical-baselines}.

\subsection{Quantum fleet power and Kardashev thresholds}

Attach a wall-plug budget to each surface-code architecture to convert fleet qubits into power.
For the superconducting point, take $12\,\mathrm{W}$ per physical qubit at $18\%$ efficiency.
For neutral atoms, take $\sim10^{-3}\,\mathrm{W}$ of circulating optical power per physical qubit at $30\%$ laser efficiency~\cite{Huft2022TrapArrays}.
For ion traps, take $3\,\mathrm{W}$ per physical qubit at $22\%$ efficiency.
Applying these budgets to the fleet sweep from \secref{sec:physical-footprint} yields an energy ladder that spans (and quickly exceeds) Kardashev Type~I/II/III thresholds.~\cite{Kardashev1964}

As a reference point, under the superconducting assumptions, the $b=32$ case (50\% success, 600-second runtime cap) draws $\sim9.5\times10^3\,\mathrm{MW}$.
At $b=64$ with the same runtime cap, power rises to $\sim4.1\times10^{13}\,\mathrm{MW}$, exceeding the $10^{16}\,\mathrm{W}$ Type~I threshold by several orders of magnitude.
Sparse partial-preimage regimes at $b=96$ reach $\sim1.7\times10^{23}\,\mathrm{MW}$, above the $10^{26}\,\mathrm{W}$ Type~II threshold.
The true-preimage regime at $b=256$ pushes far beyond any astrophysical Type~III scale.

\begin{figure*}[t]
  \centering
  \resizebox{0.92\textwidth}{!}{% Classical vs quantum energy budgets across Kardashev thresholds
\begin{tikzpicture}
\begin{axis}[
  width=13.5cm,
  height=9.5cm,
  xlabel={\textbf{Difficulty (bits)}},
  ylabel={\textbf{Power demand} $P$ \textbf{(W)}},
  xmin=28, xmax=168,
  xtick={32,64,96,128,160},
  xmajorgrids=true,
  ymode=log,
  log basis y=10,
  ymin=1e-4, ymax=1e50,
  ytick={1e0,1e8,1e16,1e24,1e32,1e40,1e48},
  yticklabels={$10^{0}$,$10^{8}$,$10^{16}$,$10^{24}$,$10^{32}$,$10^{40}$,$10^{48}$},
  minor ytick={1e-4,1e4,1e12,1e20,1e28,1e36,1e44},
  grid style={dashed, gray!25},
  ymajorgrids=true,
  yminorgrids=false,
  legend style={
    draw=gray!50,
    font=\footnotesize,
    legend columns=4,
    at={(0.5,-0.15)},
    anchor=north,
    column sep=1em,
    nodes={rounded corners=2pt, inner sep=3pt}
  },
  legend cell align=left,
  clip=false
]
  % Kardashev zone shading
  \fill[red!8, opacity=0.25] (axis cs:28,1e-4) rectangle (axis cs:168,1e16);
  \fill[red!15, opacity=0.20] (axis cs:28,1e16) rectangle (axis cs:168,1e26);
  \fill[red!22, opacity=0.18] (axis cs:28,1e26) rectangle (axis cs:168,1e36);

  % Kardashev threshold lines
  \addplot[very thick, color=red!65!black, dashed, forget plot] coordinates {(28,1e16) (168,1e16)};
  \addplot[very thick, color=red!75!black, dashed, forget plot] coordinates {(28,1e26) (168,1e26)};
  \addplot[very thick, color=red!85!black, dotted, forget plot] coordinates {(28,1e36) (168,1e36)};

  % Mainnet reference point (2025-01-01): D≈1.10×10^{14} → b≈78.6 leading-zero bits.
  \addplot[very thick, color=black, dashed, forget plot] coordinates {(78.64,1e-4) (78.64,1e50)};

  % Zone labels on right edge
  \node[anchor=east, font=\scriptsize\bfseries, text=red!60!black] at (axis cs:168,3e2) {Sub-Type~I};
  \node[anchor=east, font=\scriptsize\bfseries, text=red!65!black] at (axis cs:168,3e20) {Type~I};
  \node[anchor=east, font=\scriptsize\bfseries, text=red!75!black] at (axis cs:168,3e30) {Type~II};
  \node[anchor=east, font=\scriptsize\bfseries, text=red!85!black] at (axis cs:168,3e42) {Type~III};

  % Classical ASIC envelope
  \addplot[draw=gray!70, thick, fill=gray!40, fill opacity=0.4] coordinates {
    (32,5.72653568e-4)
    (64,2.459528346497712e6)
    (96,1.056359381179263e16)
    (128,4.5370289949877325e25)
    (160,1.948639115447606e35)
    (160,4.262648065041638e34)
    (128,9.924750926535666e24)
    (96,2.310786146329638e15)
    (64,5.380218257963746e5)
    (32,1.25267968e-4)
  } -- cycle;
  \addlegendentry{ASIC envelope}

  % Superconducting fleet
  \addplot[only marks, mark=o, mark size=4pt, line width=1.2pt, color=blue!90!black] coordinates {
    (32,9.519869333333334e9)
    (64,4.056424845985803e19)
    (96,1.742221205217716e29)
    (128,7.482783098807794e38)
    (160,3.213830869244101e48)
  };
  \addlegendentry{Supercond.}

  % Neutral atom fleet
  \addplot[only marks, mark=square, mark size=3.5pt, line width=1.2pt, color=orange!90!black] coordinates {
    (32,1.083068083333333e6)
    (64,4.64842427860193e15)
    (96,1.996483025452237e25)
    (128,8.574829301336493e34)
    (160,3.682861141802277e44)
  };
  \addlegendentry{Neutral atom}

  % Ion-trap fleet
  \addplot[only marks, mark=triangle, mark size=5pt, line width=1.2pt, color=teal!80!black] coordinates {
    (32,4.406328032727273e10)
    (64,1.892474769223248e20)
    (96,8.128117242318897e29)
    (128,3.490999773381337e39)
    (160,1.499372985701625e49)
  };
  \addlegendentry{Ion-trap}

\end{axis}
\end{tikzpicture}}
  \caption{Classical network power (solid/segmented curves) versus quantum fleet demand (markers) across difficulty bits. Horizontal bands mark Kardashev Type~I/II/III thresholds.  A dashed vertical marker highlights the Bitcoin mainnet difficulty on 2025-01-01 ($D\approx 1.1\times10^{14}$, $b\approx78.6$)~\cite{BlockchainComDifficulty,BlockchainComChartsAPI}.}
  \label{fig:kardashev-energy-budget}
\end{figure*}
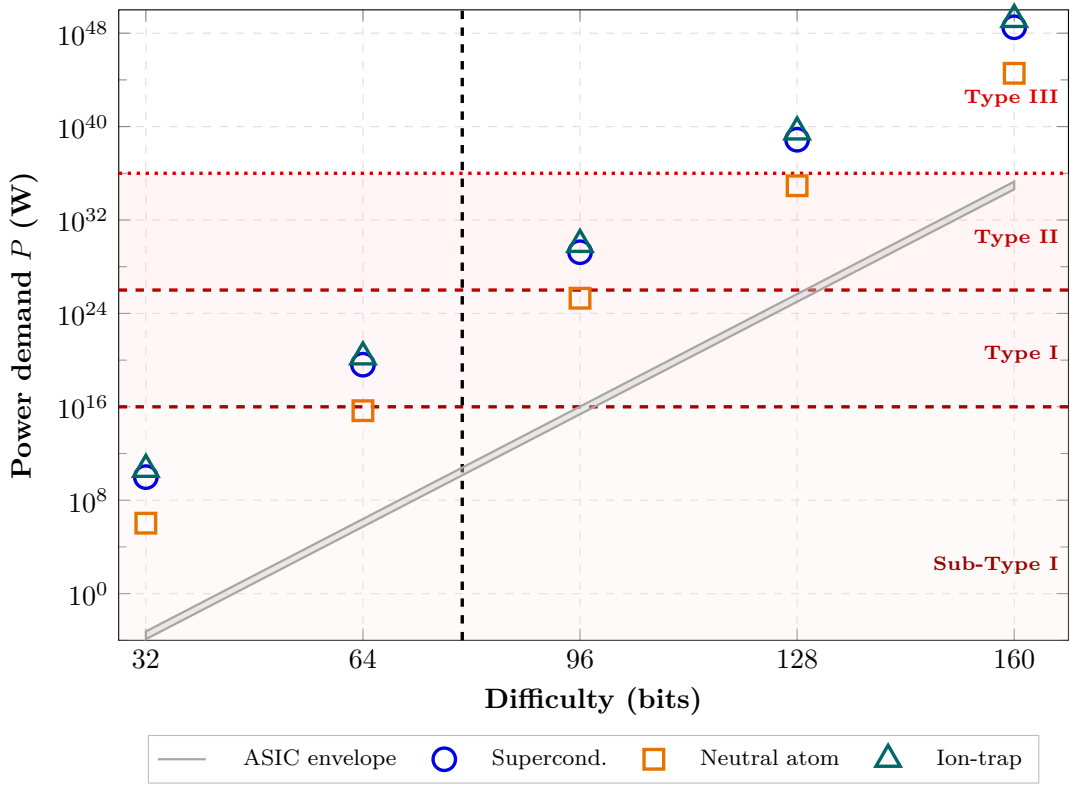

Figure~\ref{fig:kardashev-energy-budget} juxtaposes classical network power and these quantum fleet demands across difficulty bits.

\section{Discussion}
\label{sec:discussion}

\subsection{Headline result: the quantum-to-classical cost ratio}

This paper does not ask whether Bitcoin as a whole resists quantum attack.
Shor's algorithm already answers the signature-layer part in the negative~\cite{DallaireDemers2025-ECDLP}.
The narrower question here concerns mining: once we price Grover mining in physical units, does it become a practical consensus threat?
For the architectures and mining models studied here, the answer is no.
The sharpest summary reduces the result to a single observable: the ratio of quantum fleet power to classical network power.
At Bitcoin's January~2025 mainnet difficulty ($b\approx79$), the classical network draws roughly $P_{\mathrm{cl}}\sim 10\text{--}15\,\mathrm{GW}$~\cite{CBECI2025}; the superconducting quantum fleet demands $P_{\mathrm{Q}}\sim 10^{25}\,\mathrm{W}$ (Table~\ref{tab:scenario-scaling}, Figures~\ref{fig:fleet-heatmap}--\ref{fig:kardashev-energy-budget}).
The ratio $P_{\mathrm{Q}}/P_{\mathrm{cl}}\sim 10^{14}$ measures how many decades of engineering improvement a quantum miner must bridge before matching the incumbent on energy alone.
Even at the most favourable partial-preimage setting ($b=32$, $2^{224}$ marked states), the quantum fleet still draws $\sim10^{4}\,\mathrm{MW}$ against a classical network that would need only microwatts at such low difficulty. The ratio therefore stays above $10^{10}$.
In every regime the estimator explores, the quantum miner pays a steeper energy bill than the classical miner it aims to displace, confirming that Grover's quadratic speedup does not survive the multiplicative overhead of fault tolerance at Bitcoin-relevant scales.

Three independent cost drivers compound to produce this gap:
\begin{enumerate}
  \item \emph{Oracle overhead.} The reversible \dSHA oracle adds $\sim\!3\times10^5$ $T$ gates per Grover iteration (Table~\ref{tab:sha-ledger}), inflating the na\"ive $2^{b/2}$ query count by $\sim\!2^{38}$ in non-Clifford volume~\cite{Amy2017-SHA}.
  \item \emph{Distillation tax.}  Each $T$ gate requires a magic-state factory occupying $1.25\,d^2$ physical qubits; at the code distances forced by $\Ttot$ (\eqnref{eq:logical-budget-tcount}), hundreds to thousands of factories run in parallel (\eqnref{eq:factory-demand}).
  \item \emph{Fleet multiplication.}  Nakamoto's ten-minute block window caps Grover depth via $r_{\mathrm{cap}}$ (\eqnref{eq:grover-rcap}), forcing the shortfall in per-machine success $P_1$ (\eqnref{eq:single-machine-success}) to be compensated by exponentially many independent machines (\eqnref{eq:fleet-machines}).
\end{enumerate}
Together, these factors push partial-preimage fleets into the million-qubit regime even at difficulty~1 (Table~\ref{tab:example-footprint}) and approach Kardashev Type~II territory ($\sim10^{26}\,\mathrm{W}$) at current mainnet difficulty (Figure~\ref{fig:kardashev-energy-budget}).
Figures~\ref{fig:fleet-heatmap} and~\ref{fig:fleet-tradeoff} show how that escalation happens.
Figure~\ref{fig:fleet-heatmap} maps the baseline geography: grey infeasible regions where one Grover iteration already misses the deadline, then a steep climb from industrial-scale fleets in the favourable partial-preimage corner to civilization-scale fleets as marked states thin out.
Figure~\ref{fig:fleet-tradeoff} shows the same verdict from another angle: runtime does not come cheap; each step toward a faster miner demands an explosion in fleet size.
Figure~\ref{fig:kardashev-energy-budget} completes the translation from qubits to infrastructure.
At that point the machine stops resembling a computer in any ordinary sense.
It reads instead like a thermodynamic object: part cryogenic refinery, part power plant, part radiator network.
Figures~\ref{fig:energy-scale-ladder}--\ref{fig:high-energy-tradeoff} then test the science-fiction escape hatch directly.
Figure~\ref{fig:energy-scale-ladder} raises the microscopic clock through THz, optical, x-ray, nuclear, QCD, electroweak, and Planck analogs.
Figures~\ref{fig:high-energy-heatmap-low} and~\ref{fig:high-energy-heatmap-high} show that each rung trims at most about a decade of fleet scale before marked-state counts take over again.
Figure~\ref{fig:high-energy-tradeoff} shows the same limit in Pareto form.
The result does not look like a near-term processor scaled up.
It looks like engineered astrophysics.

\subsection{Implications for Nakamoto consensus}

The resource model reframes Bitcoin's ``sound money'' claim as a deadline-constrained attack cost: reversing a payment requires mining an alternative chain that outpaces the honest chain within the propagation-limited window $t_{\mathrm{cap}}$, and the fleet sizes in \secref{sec:physical-footprint}--\secref{sec:energy-kardashev} quantify the resulting physical and economic barrier.
The grey cells in Figure~\ref{fig:fleet-heatmap} mark regimes where even a single Grover iteration exceeds the wall-clock cap. Together with the steep Pareto fronts in Figure~\ref{fig:fleet-tradeoff}, they confirm that cutting runtime by an order of magnitude costs far more than an order of magnitude in qubits.

These footprints also imply that Grover-enabled miners cannot be covert entrants any time soon.  A miner fielding multi-million-qubit surface-code arrays would announce its presence through capital expenditure, energy consumption, and the atypical mining cadence of stop-and-measure strategies~\cite{Sattath2020}.
Sattath's fork-pressure mechanism and Bailey--Sattath's difficulty-increase strategy stress distinct consensus layers~\cite{Sattath2020,BaileySattath2024}. In the first, a quantum miner collapses its state upon hearing a rival block, which raises fork risk. In the second, a quantum miner manipulates timestamps to inflate epoch-wide difficulty for classical competitors. Park and Spooner press on a third lever: they tie proof validity to a timed random beacon, which cuts off the long coherent search interval that feeds quantum superlinearity, though the construction shifts the burden to beacon deployment and consensus integration~\cite{ParkSpooner2022Superlinearity}.
Not every quantum pressure channel demands majority hash power. Velner, Teutsch, and Luu show that smart-contract mining pools admit block-withholding bribes~\cite{Velner2017}; a leased quantum accelerator could matter here without dominating Bitcoin's global hashrate if it pushes hashing speed high enough, long enough, to shift the bribe threshold.
Our resource model prices a complementary bottleneck: both strategies still require extremely fast, fault-tolerant Grover oracles, and the fleet and power scales in \secref{sec:physical-footprint}--\secref{sec:energy-kardashev} keep that hardware outside near-term deployment regimes.
In the language of Benkoczi et al.~and Nerem--Gaur, our estimator instantiates the missing physical quantities: the cost per Grover step, the query rate, and the runtime cap that shapes when a miner should measure~\cite{BenkocziEtAl2022QuantumBitcoinMining,NeremGaur2023AdvantageousQuantumMining}.
Their algorithmic workflow and economic criteria remain conceptually sound, but our logical-to-physical lift drives the required \(Q\) and \(r\) so far from classical hashing that the advantageous regime recedes beyond plausible fault-tolerant deployments.
Network security therefore hinges less on a sudden 51\% attack and more on gradual shifts: if industrial actors adopt quantum accelerators, protocol designers must reassess difficulty retargeting, share validation, and fork-resolution policies~\cite{Nakamoto2008,Sattath2020,BaileySattath2024}.

\subsection{Limitations and caveats}

Several modelling choices make the estimates \emph{optimistic} lower bounds.
The header-only baseline omits the reversible Merkle-root recomputation that a protocol-faithful search would require (Table~\ref{tab:mining-knob-sensitivity}).
The $T$-count failure budget (\eqnref{eq:logical-budget-tcount}) is less conservative than the spacetime-volume proxy, though Table~\ref{tab:failure-budget-sensitivity} shows the latter inflates footprints by at most $1.8\times$.
The per-qubit power band ($10^{-3}$--$12\,\mathrm{W}$) spans four orders of magnitude between Fellous-Asiani's optimistic floor and Parker--Vermeer's empirical extrapolation~\cite{FellousAsiani2023,ParkerVermeer2023}.
Even the bottom of that band cannot rescue the true-preimage fleet, whose $\sim10^{75}$ qubits would exceed any astrophysical power source.

\subsection{Future directions}

Improved reversible circuits for \SHA~\cite{Cuccaro2004,Gidney2018Adder,Maslov2016RelativePhaseToffoli} or alternative codes with cheaper non-Clifford supply could relax the resource burden, though Babbush~\emph{et~al.} caution that quadratic speedups rarely cross the practicality bar without dramatic hardware advances~\cite{Babbush2021,Eisert2025MindTheGaps}, and Cade~\emph{et~al.}'s non-asymptotic Grover analysis confirms that constant prefactors keep the crossover far from present-day device scales~\cite{Cade2023QuantifiedGrover}.
Extending the estimator to bivariate-bicycle low-density parity-check (LDPC) codes or bosonic cat codes would test whether a qualitatively different error-correction architecture can close the gap. Both promise lower non-Clifford overhead than the surface code.
Coupling energy and cooling models with economic simulations of difficulty-retargeting dynamics would sharpen Kardashev-style forecasts.
Another direction drops the majority-hash-power frame and asks when a leased quantum accelerator could distort pool incentives instead. Velner, Teutsch, and Luu show that smart-contract pools admit block-withholding bribes~\cite{Velner2017}; pricing the lease cost and transient hash-rate boost needed to make those bribes pay would test whether ASIC economics still suppress that attack in practice.
An orthogonal direction replaces Grover search entirely: sampling-based proof-of-work schemes that anchor mining to the hardness of simulating quantum optical devices sidestep the fault-tolerance overhead analysed here, though they introduce their own spoofing and verification challenges~\cite{Singh2025PoWQuantumSampling}.

\section*{Acknowledgements}
% (Optional) Funding, collaborations, discussions.
We acknowledge support from BTQ Technologies through a research grant.  We thank Gavin Brennen, Chris Tam, and Gopi Muraleedharam for interesting discussions.

% -------------------- Bibliography ----------------------
% Use natbib + Quantum's bst. Ensure your .bib entries include DOIs.

\clearpage
\section*{Symbol glossary}
\label{sec:symbol-glossary}

\begingroup
\footnotesize
\setlength{\tabcolsep}{4pt}
\renewcommand{\arraystretch}{1.15}
\begin{longtable}{@{}p{0.22\linewidth}>{\raggedright\arraybackslash}p{\dimexpr0.78\linewidth-2\tabcolsep\relax}@{}}
  \toprule
  Symbol & Meaning \\
  \midrule
\endfirsthead

  \toprule
  Symbol & Meaning \\
  \midrule
\endhead

  \bottomrule
\endlastfoot
  $H(\cdot)$ & Mining hash function, $H=\dSHA$ on the block header; treat the output as uniform over $\{0,1\}^{256}$ when estimating marked-state density. \\
  $m$ & Candidate block header (search input) supplied to $H$. \\
  \texttt{extraNonce} & Miner-controlled field in the coinbase transaction used to vary the Merkle root and extend the search beyond the 32-bit nonce. \\
  \texttt{merkle\_root} & 256-bit header field that commits to the transaction set; changing the coinbase or transaction set changes this field. \\
  $T$ & Proof-of-work target threshold in the predicate $H(m)<T$. \\
  $T_1$ & Difficulty-1 target. \\
  $D$ & Difficulty, defined by $T=T_1/D$. \\
  $p$ & Marked fraction / one-shot hit probability, $p=\Pr[H(m)<T]=T/2^{256}$. \\
  $b$ & Difficulty bits, $b:=-\log_2 p$; a power-of-two target $T=2^{256-b}$ makes $b$ match the leading-zero bit count in $H(m)$. \\
  $t_{\mathrm{blk}}$ & Mean inter-block time target (set to $600\,\mathrm{s}$ for Bitcoin). \\
  $n$ & Search-register size in bits/qubits. \\
  $N$ & Search-space size, $N=2^n$. \\
  $M$ & Marked inputs, $M=pN$. \\
  $\theta$ & Grover rotation angle, $\theta=\arcsin\sqrt{M/N}=\arcsin\sqrt{p}$. \\
  $r$ & Grover iteration count, Eq.~\eqref{eq:grover-iterations}; for $p\ll1$, $r\approx \pi/(4\sqrt{p})$. \\
  $k_{\mathrm{hash}}$ & Hash-work multiplier accounting for Merkle-root recomputation, Eq.~\eqref{eq:khash-merkle}. \\
  $\alpha_{\mathrm{merkle}}$ & Toggle for coherent Merkle-root recomputation: 0 (header-only) or 1 (protocol-faithful). \\
  $\beta_{\mathrm{midstate}}$ & Midstate-reuse factor: 1 (no reuse) or $1/2$ (nonce-only mining with midstate). \\
  $t_{\mathrm{iter}}$ & Wall-clock time for one Grover iteration on one machine. \\
  $t_{\mathrm{cap}}$ & Per-machine wall-clock runtime cap for a mining run. \\
  $r_{\mathrm{cap}}$ & Time-capped Grover iteration count, Eq.~\eqref{eq:grover-rcap}. \\
  $P_1$ & Single-machine success probability under $r_{\mathrm{cap}}$, Eq.~\eqref{eq:single-machine-success}. \\
  $P_t$ & Target end-to-end success probability for a mining run. \\
  $N_{\mathrm{machines}}$ & Fleet size needed to reach $P_t$ given $P_1$, Eq.~\eqref{eq:fleet-machines}. \\
  $Q_{\mathrm{machine}}$ & Physical-qubit footprint of one machine (data patches plus factories). \\
  $Q_{\mathrm{fleet}}$ & Total physical qubits across the fleet, $Q_{\mathrm{fleet}}=N_{\mathrm{machines}}Q_{\mathrm{machine}}$. \\
	  $\Toracle$ & $T$-count per Grover iteration, Eq.~\eqref{eq:tcount-grover}. \\
	  $\Tdiff$ & Diffusion-reflection $T$-count overhead for an $n$-qubit register. \\
	  $\Ttot$ & Total $T$-count for a run, $\Ttot=r\times\Toracle$. \\
	  $p_{\mathrm{run}}$ & Run-level logical-failure budget allocated to fault-tolerant execution (set to $0.01$ in \secref{sec:physical-footprint}). \\
	  $p_{\mathrm{fail}}$ & Per-location logical-failure target, $p_{\mathrm{fail}}=0.01/\Ttot$, Eq.~\eqref{eq:logical-budget-tcount}. \\
	  $p_{\mathrm{phys}}$ & Physical two-qubit error rate. \\
  $p_L$ & Logical error rate per code cycle for a distance-$d$ patch, Eq.~\eqref{eq:surface-logical}. \\
  $d$ & Surface-code distance. \\
  $\tau$ & Code-cycle time. \\
  $t_{\mathrm{logical}}$ & Logical runtime of a Grover run, obtained by multiplying the program depth in code cycles by $\tau$. \\
  $V$ & Logical spacetime volume proxy, $V=Q_{\mathrm{tot}}(t_{\mathrm{logical}}/\tau)$, used in Table~\ref{tab:failure-budget-sensitivity}. \\
  $R_T$ & Required $T$-state throughput, Eq.~\eqref{eq:factory-demand}. \\
  $N_{\mathrm{fac}}$ & Number of $T$-state factories sized to supply $R_T$. \\
  $Q_{\mathrm{tot}}$ & Logical-qubit footprint of the Grover program (search register plus oracle workspace). \\
\end{longtable}
\endgroup

\clearpage

\appendix
\section{Reversible circuit modules for \SHA}
\label{app:rev-sha}
% Specify the reversible/clean circuit construction for \SHA:
% - Message schedule, compression function, round constants.
% - Choice of adder (Cuccaro, Draper, etc.), XOR network synthesis, rotation/shift gates.
% - Ancilla management / uncomputation strategy.
% - Circuit depth, T-count/T-depth (or Toffoli count) per hash evaluation.
Build the \SHA oracle as a clean machine: compute, use, uncompute. Three habits guide every line:
(i) treat linear mix as CNOT networks;
(ii) rename wires for rotations, write fresh for shifts;
(iii) reckon addition as the driver of the \(T\)-budget.

% ---------- Reversible-update notation ----------
\providecommand{\xorup}{\mathrel{\oplus=}}     % in-place XOR update
\providecommand{\addup}{\mathrel{\boxplus=}}   % in-place mod-2^32 add update
\newcommand{\ROTR}{\mathrm{ROTR}}
\newcommand{\ROTL}{\mathrm{ROTL}}
\newcommand{\SHR}{\mathrm{SHR}}
% All word operations below act bitwise on 32 lanes unless stated otherwise.

% ---------- Primitive reversible circuits ----------
\paragraph{Linear / XOR layer.}
Compose any XOR-only layer from CNOTs; synthesize the network by row-reduction \cite{Patel2008LinearReversible}.

\paragraph{Rotate and shift.}
Rotate right; permute wires; relabel only---no gates.
Shift right; write into a fresh 32-bit word; uncompute after use.

\paragraph{Big / small sigma layers (write into a destination word).}
\begin{align}
S_0 &:= \Sigma_0(a) \ =\ \ROTR^{2}(a)\ \oplus\ \ROTR^{13}(a)\ \oplus\ \ROTR^{22}(a),
\label{eq:rev-sigma0} \\
S_1 &:= \Sigma_1(e) \ =\ \ROTR^{6}(e)\ \oplus\ \ROTR^{11}(e)\ \oplus\ \ROTR^{25}(e),
\label{eq:rev-sigma1} \\
s_0(x) &:= \sigma_0(x)\ =\ \ROTR^{7}(x)\ \oplus\ \ROTR^{18}(x)\ \oplus\ \SHR^{3}(x),
\label{eq:rev-sigma-small0} \\
s_1(x) &:= \sigma_1(x)\ =\ \ROTR^{17}(x)\ \oplus\ \ROTR^{19}(x)\ \oplus\ \SHR^{10}(x).
\label{eq:rev-sigma-small1}
\end{align}

\paragraph{Choice (\texorpdfstring{$\mathrm{Ch}$}{Ch}) into a destination word.}
Exploit
\begin{equation}
\mathrm{Ch}(x,y,z)\ =\ z\ \oplus\ \bigl(x\land (y\oplus z)\bigr).
\label{eq:rev-ch-id}
\end{equation}
Use one 32-bit scratch \(T\) and one Toffoli per lane (relative-phase Toffoli trims \(T\)-count \cite{Maslov2016RelativePhaseToffoli}):
\begin{align}
C &\leftarrow z, \label{eq:rev-ch-seq-1}\\
T &\leftarrow y \oplus z, \label{eq:rev-ch-seq-2}\\
C &\xorup x \land T, \label{eq:rev-ch-seq-3}\\
T &\leftarrow T \oplus y \oplus z \ =\ 0^{32}. \label{eq:rev-ch-seq-4}
\end{align}

\paragraph{Majority (\texorpdfstring{$\mathrm{Maj}$}{Maj}) into a destination word.}
Use
\begin{equation}
\mathrm{Maj}(x,y,z)\ =\ (x\land y)\ \oplus\ \bigl(z\land (x\oplus y)\bigr).
\label{eq:rev-maj-id}
\end{equation}
Keep one 32-bit scratch \(T\); spend two Toffolis per lane:
\begin{align}
M &\leftarrow 0^{32}, \label{eq:rev-maj-seq-1}\\
M &\xorup x \land y, \label{eq:rev-maj-seq-2}\\
T &\leftarrow x \oplus y, \label{eq:rev-maj-seq-3}\\
M &\xorup z \land T, \label{eq:rev-maj-seq-4}\\
T &\leftarrow T \oplus x \oplus y \ =\ 0^{32}. \label{eq:rev-maj-seq-5}
\end{align}

\paragraph{32-bit modular adder (CDKM ripple \cite{Cuccaro2004}).}
Use the adder as a black box with one clean carry ancilla \(c\):
\begin{equation}
\mathrm{Add}_{32}(x,y;c):\ (x,y,c{=}0)\ \mapsto\ \bigl(x,\ x\boxplus y,\ c{=}0\bigr).
\label{eq:rev-adder-spec}
\end{equation}
Gate and width ledger (standard form):
\begin{equation}
\mathrm{Toffoli}(\mathrm{Add}_{n}) = 2n-1,\quad
\mathrm{CNOT}(\mathrm{Add}_{n}) = 5n-3,\quad
\text{ancilla} = 1.
\label{eq:rev-adder-cost}
\end{equation}
Gidney's measurement-assisted AND drives the \(T\)-count for an \(n\)-bit add to \(\approx 4n\) \cite{Gidney2018Adder}.

\paragraph{Constant adder.}
\(\mathrm{AddConst}_{32}(K;x;c):\ (x,c{=}0)\mapsto (x\boxplus K,c{=}0)\) with fewer controls on average \cite{Oumarou2022ConstAdder}.

% ---------- Register layout ----------
\paragraph{Registers.}
Hold state words \(a,b,c,d,e,f,g,h\in\{0,1\}^{32}\).
Keep a 16-word message ring \(W[0..15]\).
Reuse two 32-bit scratch words \(S,T\) across all subroutines, plus one clean carry ancilla \(c\).

\paragraph{Total width (baseline plus Grover overhead).}
\begin{equation}
Q_{\mathrm{tot}}\ =\ 8\cdot 32\ +\ 16\cdot 32\ +\ 2\cdot 32\ +\ 1\ +\ n\ +\ Q_{\mathrm{cmp}}\ +\ Q_{\mathrm{diff}}\ =\ 833\ +\ n\ +\ Q_{\mathrm{cmp}}\ +\ Q_{\mathrm{diff}}\ \text{qubits}.
\label{eq:rev-width}
\end{equation}

Here $n$ accounts for the Grover search register, while $Q_{\mathrm{cmp}}$ and $Q_{\mathrm{diff}}$ capture the ancillae used by the comparator and diffusion operators; equivalently, one may view this as updating the baseline width via $Q_{\mathrm{tot}} \leftarrow Q_{\mathrm{tot}} + n + Q_{\mathrm{cmp}} + Q_{\mathrm{diff}}$.

% ---------- Reversible message schedule ----------
\paragraph{Message schedule (ring buffer, three adds for $t\ge 16$).}
Reuse \(T\) as the accumulator; keep only \(S\) and \(T\) as scratch:
\begin{align}
S &\leftarrow s_1\!\bigl(W_{t-2}\bigr), \label{eq:rev-Wt-1}\\
T &\leftarrow S\ \boxplus\ W_{t-7} \quad \text{via }\mathrm{Add}_{32}, \label{eq:rev-Wt-2}\\
S &\leftarrow s_0\!\bigl(W_{t-15}\bigr), \label{eq:rev-Wt-3}\\
T &\leftarrow T\ \boxplus\ S \quad \text{via }\mathrm{Add}_{32}, \label{eq:rev-Wt-4}\\
W_t &\leftarrow T\ \boxplus\ W_{t-16} \quad \text{into slot }(t\bmod 16), \label{eq:rev-Wt-5}\\
S &\leftarrow 0^{32}. \label{eq:rev-Wt-6}
\end{align}
For \(t\le 15\), load \(W_t\) from the parsed 512-bit message block~\cite{FIPS180-4}.

% ---------- One reversible compression round ----------
\paragraph{Round $t$ (state path uses seven adds; steady state uses ten adds including the schedule).}
\begin{align}
S_1 &\leftarrow \Sigma_1(e), \label{eq:rev-rnd-1}\\
C   &\leftarrow \mathrm{Ch}(e,f,g) \quad \text{via }\eqref{eq:rev-ch-seq-1} \text{--} \eqref{eq:rev-ch-seq-4}, \label{eq:rev-rnd-2}\\
t_1 &\leftarrow h\ \boxplus\ S_1 \quad \text{via }\mathrm{Add}_{32}, \label{eq:rev-rnd-3}\\
t_1 &\leftarrow t_1\ \boxplus\ C \quad \text{via }\mathrm{Add}_{32}, \label{eq:rev-rnd-4}\\
t_1 &\leftarrow t_1\ \boxplus\ K_t \quad \text{via }\mathrm{AddConst}_{32}, \label{eq:rev-rnd-5}\\
T_1 &\leftarrow t_1\ \boxplus\ W_t \quad \text{via }\mathrm{Add}_{32}, \label{eq:rev-rnd-6}\\[2mm]
S_0 &\leftarrow \Sigma_0(a), \label{eq:rev-rnd-7}\\
M   &\leftarrow \mathrm{Maj}(a,b,c) \quad \text{via }\eqref{eq:rev-maj-seq-1} \text{--} \eqref{eq:rev-maj-seq-5}, \label{eq:rev-rnd-8}\\
T_2 &\leftarrow S_0\ \boxplus\ M \quad \text{via }\mathrm{Add}_{32}, \label{eq:rev-rnd-9}\\[2mm]
e   &\leftarrow d\ \boxplus\ T_1 \quad \text{via }\mathrm{Add}_{32}, \label{eq:rev-rnd-10}\\
a   &\leftarrow T_1\ \boxplus\ T_2 \quad \text{via }\mathrm{Add}_{32}, \label{eq:rev-rnd-11}\\
(b,c,d,f,g,h) &\leftarrow (a_{\text{old}},\,b_{\text{old}},\,c_{\text{old}},\,e_{\text{old}},\,f_{\text{old}},\,g_{\text{old}}). \label{eq:rev-rnd-12}
\end{align}
Finally uncompute temporaries: \(S_1\!\leftarrow 0^{32}\), \(S_0\!\leftarrow 0^{32}\), \(C\!\leftarrow 0^{32}\), \(M\!\leftarrow 0^{32}\).

% ---------- Feed-forward ----------
\paragraph{Feed-forward (after 64 rounds).}
\begin{equation}
\bigl(H_0^{(j+1)},\dots,H_7^{(j+1)}\bigr)\ \leftarrow\
\bigl(H_0^{(j)}\boxplus a,\ H_1^{(j)}\boxplus b,\ \dots,\ H_7^{(j)}\boxplus h\bigr),
\label{eq:rev-feed-forward}
\end{equation}
with eight calls to \(\mathrm{Add}_{32}\), matching the FIPS~180-4 feed-forward step~\cite{FIPS180-4}.

% ---------- Quick resource ledger ----------
\paragraph{Round-level ledger (word width $=32$).}
Reckon adders per round and Toffolis from \(\mathrm{Ch}\) and \(\mathrm{Maj}\):
\begin{align}
A_{\text{state}} &= 7, &
A_{\text{steady}} &= 10, \label{eq:rev-A-counts}\\
T_{\mathrm{bool}} &= 1\cdot 32\ +\ 2\cdot 32\ =\ 96.
\label{eq:rev-Toffoli-bool}
\end{align}
With Gidney adders (\(\approx 4n\) \(T\)-gates per \(n\)-bit add \cite{Gidney2018Adder}) and relative-phase Toffoli (4 \(T\)-gates \cite{Maslov2016RelativePhaseToffoli}), an order-of-magnitude \(T\)-count per round reads
\begin{equation}
T_{\text{round}}\ \approx\ 128\,A\ +\ 4\,T_{\mathrm{bool}}
\ \in\
\begin{cases}
[\,1280,\ 1664\,] & \text{for } A\in\{7,10\}.
\end{cases}
\label{eq:rev-T-round}
\end{equation}

% ---------- Quantikz schematic ----------
\begin{figure}[t]
\centering
\ifhasquantikz
  \begin{quantikz}[row sep=0.9ex, column sep=1.2ex]
  % Maj path on (a,b,c)
  \lstick{$a$} & \gate{\Sigma_0} & \gate[wires=3]{\mathrm{Maj}} & \qw & \qw & \qw \\
  \lstick{$b$} & \qw             & \ghost{\mathrm{Maj}}          & \qw & \qw & \qw \\
  \lstick{$c$} & \qw             & \ghost{\mathrm{Maj}}          & \qw & \qw & \qw \\
  % d just shifts
  \lstick{$d$} & \qw             & \qw                           & \qw & \qw & \qw \\
  % Ch path on (e,f,g) feeding T1 adds
  \lstick{$e$} & \gate{\Sigma_1} & \qw                           & \qw        & \qw        & \qw \\
  \lstick{$f$} & \qw             & \gate[wires=3]{\mathrm{Ch}}   & \qw        & \qw        & \qw \\
  \lstick{$g$} & \qw             & \ghost{\mathrm{Ch}}           & \qw        & \qw        & \qw \\
  \lstick{$h$} & \qw             & \ghost{\mathrm{Ch}}           & \gate{+\!} & \qw        & \qw \\
  % constants / schedule added into T1
  \lstick{$K_t$} & \qw           & \qw                           & \gate{+\!} & \qw        & \qw \\
  \lstick{$W_t$} & \qw           & \qw                           & \gate{+\!} & \qw        & \qw \\
  % combine with T2 (Maj+Sigma0) conceptually via +
  \lstick{}      & \qw           & \qw                           & \qw        & \gate{+\!} & \qw
  \end{quantikz}
\else
  \fbox{\parbox{0.82\linewidth}{\centering
      Quantum circuit diagram omitted. Compile with the \texttt{quantikz} package installed to display the full reversible round schematic.}}
\fi
\caption{One reversible round, drawn for intuition: \(\Sigma\) blocks use XOR only; \(\mathrm{Ch}\) spans \((e,f,g)\); \(\mathrm{Maj}\) spans \((a,b,c)\); “\(+\)” marks 32-bit adders.}
\label{fig:sha256-round}
\end{figure}
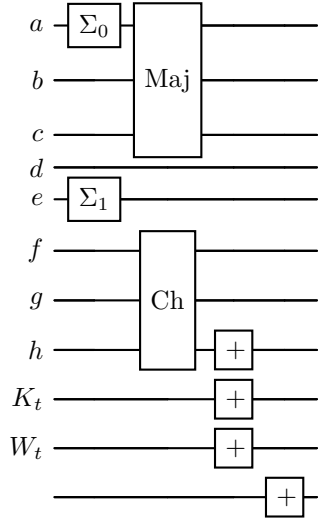

% ---------- Double SHA-256 wrapper ----------
\paragraph{Double-hash oracle.}
\begin{equation}
(m,\ 0^{256},\ 0^{256})\
\longmapsto\
\bigl(m,\ \SHA(m),\ \SHA(\SHA(m))\bigr),
\label{eq:rev-dsha-wrapper}
\end{equation}
with full uncomputation between the two calls, implementing
\[
\dSHA(m):=\SHA(\SHA(m))
\]
as specified by FIPS~180-4~\cite{FIPS180-4,Nakamoto2008}.

\subsection{Reversible circuit for \RIPEMD}
\label{app:rev-ripemd}

\RIPEMD mirrors \SHA's reversible habits but walks two parallel five-word branches.  Each branch maintains registers $(A,B,C,D,E)\in\{0,1\}^{32}$ and the classical permutation of the $16$-word message ring.  Rotations remain wire relabelings; shifts write into fresh words and uncompute once consumed.  Two shared scratch words and a single clean carry ancilla suffice for the CDKM adders.

\paragraph{Boolean layer.}
Each round uses one of the five nonlinear functions
\begin{align*}
f_0(x,y,z) &= x \oplus y \oplus z, & f_1(x,y,z) &= (x\land y) \lor (\bar{x} \land z), \\
f_2(x,y,z) &= (x \lor \bar{y}) \oplus z, & f_3(x,y,z) &= (x \land z) \lor (y \land \bar{z}), \\
f_4(x,y,z) &= x \oplus (y \lor \bar{z}).
\end{align*}
The Clifford-only $f_0$ applies in the first and last $16$ rounds across the two branches; the remaining $64$ rounds each incur a single relative-phase Toffoli per bit lane, or $32$ Toffolis per round.

\paragraph{Round structure.}
For round index $j$ the reversible update follows the classical specification:
\begin{align*}
T &:= \ROTL_{s_j}\bigl(A + f_j(B,C,D) + X_{r_j} + K_j\bigr),\\
(A,B,C,D,E) &\leftarrow (E, T, \ROTL_{10}(B), C, D).
\end{align*}
We count four 32-bit additions per round (three data-dependent, one constant) per branch and reuse the same adder template for both paths.

\paragraph{Feed-forward.}
After $80$ rounds per branch, the two paths recombine with the incoming chaining variables via five pairwise sums, consuming ten additional additions.

\paragraph{Ledger.}
Summing the nonlinear layers and adders over both branches produces the ledger in Table~\ref{tab:ripemd-ledger}.  Our reference implementation reproduces the classical digest on standard test vectors while matching these counts exactly.  Replacing the relative-phase Toffoli scaffolds with standard CCX synthesis adds three $T$ gates per Toffoli; the table annotates the resulting T-count penalty.

\begin{table}[t]
  \centering
  \caption{Logical ledger for one forward evaluation of \RIPEMD.  ``Adders'' counts $32$-bit CDKM adders; ``boolean'' counts relative-phase Toffolis from the non-linear layer.  T-count entries list the relative-phase tally with the additional standard-Toffoli penalty in parentheses.}
  \label{tab:ripemd-ledger}
  \scriptsize
  \setlength{\tabcolsep}{3pt}
  \renewcommand{\arraystretch}{1.15}
  \begin{tabularx}{\linewidth}{>{\raggedright\arraybackslash}XYYYYY}
    \toprule
    Stage & Adders & Boolean Toffolis & Total Toffolis & T-count & CNOTs \\
    \midrule
    Parallel rounds (2\,$\times$80) & \num{640} & \num{4096} & \num{44416} & \makecell[r]{\num{98304}\\{\scriptsize$+\num{133248}\,\text{std}$}} & \num{108672} \\
    Feed-forward mix & \num{10} & \num{0} & \num{630} & \makecell[r]{\num{1280}\\{\scriptsize$+\num{1890}\,\text{std}$}} & \num{1570} \\
    \midrule
    \textbf{Forward hash} & \num{650} & \num{4096} & \num{45046} & \makecell[r]{\num{99584}\\(+\num{135138}\,\text{std})} & \num{110242} \\
    \bottomrule
  \end{tabularx}
  \renewcommand{\arraystretch}{1.0}
\end{table}

The reversible design spans $897$ logical qubits: two five-word branches ($2\times5\times32$), the $16$-word schedule ($512$ qubits), two shared scratch words, and one ancilla.

% ECDLP ladder moved to ARCHIVE_removed_appendices.tex
% Appendices moved to standalone arXiv documents:
% \input{tex/appendix_application_landscape}
% \input{tex/appendix_simulation_yardstick}
% \input{tex/appendix_boson_sampling_simulation}
% \input{tex/appendix_boson_sampling_ftqc}
\section{Scale-invariant principles and high-energy surface codes}
\label{app:scale-invariant-hesc}

\paragraph{Principles.}
The operational content of quantum information processing reads as \emph{scale-agnostic}: (i) quantum Turing machines and uniform quantum circuit families simulate each other with at most polynomial overhead.~\cite{Yao1993QuantumCircuitComplexity,BernsteinVazirani1997QuantumComplexity} (ii) Anyonic/topological models admit fault tolerance and computational universality (hence equivalence to the circuit model), with locality and threshold properties carried by codes such as the toric/surface code.~\cite{Kitaev2003AnyonsFTQC,Dennis2002TopologicalMemory,BravyiKitaev1998LatticeBoundaries,FreedmanKitaevLarsenWang2003TQC,FreedmanLarsenWang2002ModularFunctor,FreedmanKitaevWang2002SimTQFT,Nayak2008NonAbelianTQC,AharonovBenOr2008Threshold}
(iii) Effective topological quantum field theories (TQFTs) abstract away metric/energy dependence.~\cite{Atiyah1988TQFT,Witten1988TQFT,Witten1989JonesPolynomial} We therefore treat ``surface code at energy scale $E$'' as the same computational primitive paced by a different microscopic \emph{clock}.

\paragraph{High-Energy Surface-Code Hypothesis (HESC).}
Assume that for a ladder of microscopic energy scales $E$ we access metastable, local, gapped condensates (``toric-code matter'') that support:
\begin{enumerate}
  \item planar/surface-code stabilizers realized by short-range interactions on a 2D lattice with boundaries;~\cite{BravyiKitaev1998LatticeBoundaries,Dennis2002TopologicalMemory}
  \item syndrome readout and feed-forward with overhead $\kappa$ that stays \emph{bounded} across the ladder (not necessarily constant);
  \item error models that keep the threshold within the same order as the microwave baseline.
\end{enumerate}
Under HESC, the code-cycle time scales as
\begin{equation}
  \tau_{\mathrm{cyc}}(E) = \kappa(E)\,\frac{h}{E}, \qquad
  \ell(E) = \frac{\hbar c}{E},
  \label{eq:tau-and-ell-from-E}
\end{equation}
so any wall-time-limited resource (for example, depth-limited Grover iterations) grows $\propto E/\kappa(E)$ when the physical time budget stays fixed. Ultimate device limits remain bounded by Margolus--Levitin/Bekenstein arguments as surveyed by Lloyd.~\cite{Lloyd1999Ultimate}

\paragraph{Energy--time--length ladder.}
Figure~\ref{fig:energy-scale-ladder} charts the HESC ladder: the left axis tracks the code-cycle time $\tau_{\mathrm{cyc}}$ and reduced wavelength floor $\ell$, while the right axis plots the Grover depth multiplier $S(E)$.  We anchor $\kappa(5~\mathrm{GHz}) = 5{\times}10^3$ so that $\tau_{\mathrm{cyc}} = 1\,\mu\mathrm{s}$, then march the condensate energy through THz phononics, optical tweezers, x-ray defects, and beyond-Standard-Model analogs.  Table~\ref{tab:hesc-energy-ladder} lists the same ladder numerically, making the trade-off between microscopic clock speed and control overhead explicit.

\begin{figure*}[t]
  \centering
  \IfFileExists{figures/energy_scale_ladder.tikz}{%
    % Energy-scale ladder figure for high-energy surface-code discussion
\begin{tikzpicture}
  \begin{axis}[
    name=mainaxis,
    width=0.92\textwidth,
    height=0.55\textwidth,
    xmode=log,
    ymode=log,
    log basis x=10,
    log basis y=10,
    xlabel={\textbf{Energy scale} $E$ \textbf{(eV)}},
    ylabel={\textbf{Cycle time} $\tau_{\mathrm{cyc}}$ \textbf{(s)} / \textbf{Length} $\ell$ \textbf{(m)}},
    axis y line*=left,
    axis x line*=bottom,
    ymin=1e-45,
    ymax=1e-1,
    xmin=1e-6,
    xmax=1e30,
    ytick={1e-42,1e-36,1e-30,1e-24,1e-18,1e-12,1e-6,1e0},
    grid=major,
    grid style={gray!25, dashed},
    tick label style={font=\scriptsize},
    label style={font=\small},
    clip=true,
    legend style={
      draw=gray!50,
      font=\footnotesize,
      legend columns=3,
      at={(0.5,-0.22)},
      anchor=north,
      column sep=1.5em,
      nodes={rounded corners=2pt, inner sep=3pt}
    },
  ]
    % Code cycle time curve
    \addplot+[color=blue!80!black, thick, mark=*, mark size=2.5pt]
      coordinates {
        (2.067834e-05,1.0e-6)
        (1.0e-02,4.136e-10)
        (2.0e+00,2.068e-13)
        (1.0e+04,4.136e-18)
        (1.0e+06,2.068e-20)
        (1.0e+08,8.272e-23)
        (1.0e+11,4.136e-26)
        (1.0e+12,4.136e-27)
        (1.0e+19,4.136e-34)
        (1.221e+28,3.387e-43)
      };
    \addlegendentry{Code cycle $\tau_{\mathrm{cyc}}$}

    % Wavelength floor curve
    \addplot+[color=orange!90!red, thick, mark=square*, mark size=2.5pt]
      coordinates {
        (2.067834e-05,9.543e-03)
        (1.0e-02,1.973e-05)
        (2.0e+00,9.866e-08)
        (1.0e+04,1.973e-11)
        (1.0e+06,1.973e-13)
        (1.0e+08,1.973e-15)
        (1.0e+11,1.973e-18)
        (1.0e+12,1.973e-19)
        (1.0e+19,1.973e-26)
        (1.221e+28,1.616e-35)
      };
    \addlegendentry{Wavelength floor $\ell$}

    % Legend-only entry for Grover (actual curve on second axis)
    \addlegendimage{color=purple!80!black, thick, mark=triangle*, mark size=3pt}
    \addlegendentry{Grover gain $S(E)$}

    % Platform annotations with staggered positioning
    \node[anchor=south west, font=\scriptsize, fill=white, fill opacity=0.9, text opacity=1, inner sep=2pt, rounded corners=1pt] 
      at (axis cs:2e-02,8e-10) {THz condensates};
    \node[anchor=south west, font=\scriptsize, fill=white, fill opacity=0.9, text opacity=1, inner sep=2pt, rounded corners=1pt] 
      at (axis cs:4e+00,4e-13) {Optical tweezers};
    \node[anchor=south west, font=\scriptsize, fill=white, fill opacity=0.9, text opacity=1, inner sep=2pt, rounded corners=1pt] 
      at (axis cs:2e+04,8e-18) {Hard x-ray defects};
    \node[anchor=south west, font=\scriptsize, fill=white, fill opacity=0.9, text opacity=1, inner sep=2pt, rounded corners=1pt] 
      at (axis cs:2e+08,2e-22) {QCD-like flux tubes};
    \node[anchor=south east, font=\scriptsize, fill=white, fill opacity=0.9, text opacity=1, inner sep=2pt, rounded corners=1pt] 
      at (axis cs:8e+27,1e-42) {Planck frontier};

  \end{axis}

  % Secondary y-axis for Grover speedup
  \begin{axis}[
    width=0.92\textwidth,
    height=0.55\textwidth,
    xmode=log,
    ymode=log,
    log basis x=10,
    log basis y=10,
    xmin=1e-6,
    xmax=1e30,
    ymin=1e-1,
    ymax=1e33,
    ytick={1e0,1e6,1e12,1e18,1e24,1e30},
    axis y line*=right,
    axis x line=none,
    ylabel={\textbf{Grover speedup} $S(E)$},
    yticklabel style={font=\scriptsize},
    tick label style={font=\scriptsize},
    label style={font=\small},
    clip=true,
  ]
    % Grover speedup curve
    \addplot+[color=purple!80!black, thick, mark=triangle*, mark size=3pt]
      coordinates {
        (2.067834e-05,1.0e+00)
        (1.0e-02,2.42e+03)
        (2.0e+00,1.0e+05)
        (1.0e+04,1.0e+08)
        (1.0e+06,4.0e+09)
        (1.0e+08,1.0e+11)
        (1.0e+11,1.0e+14)
        (1.0e+12,1.0e+15)
        (1.0e+19,1.0e+22)
        (1.221e+28,1.22e+31)
      };
  \end{axis}
\end{tikzpicture}%
  }{%
    \fbox{\parbox{0.9\linewidth}{\centering Energy-scale ladder figure missing.}}%
  }%
  \caption{Energy ladder for the high-energy surface-code hypothesis.  Blue circles show the cycle time $\tau_{\mathrm{cyc}}$, orange squares give the reduced wavelength floor $\ell$, and purple triangles plot the Grover speedup $S(E)$ as the microscopic energy increases.  Annotations mark speculative platforms that keep locality while shrinking the clock.}
  \label{fig:energy-scale-ladder}
\end{figure*}
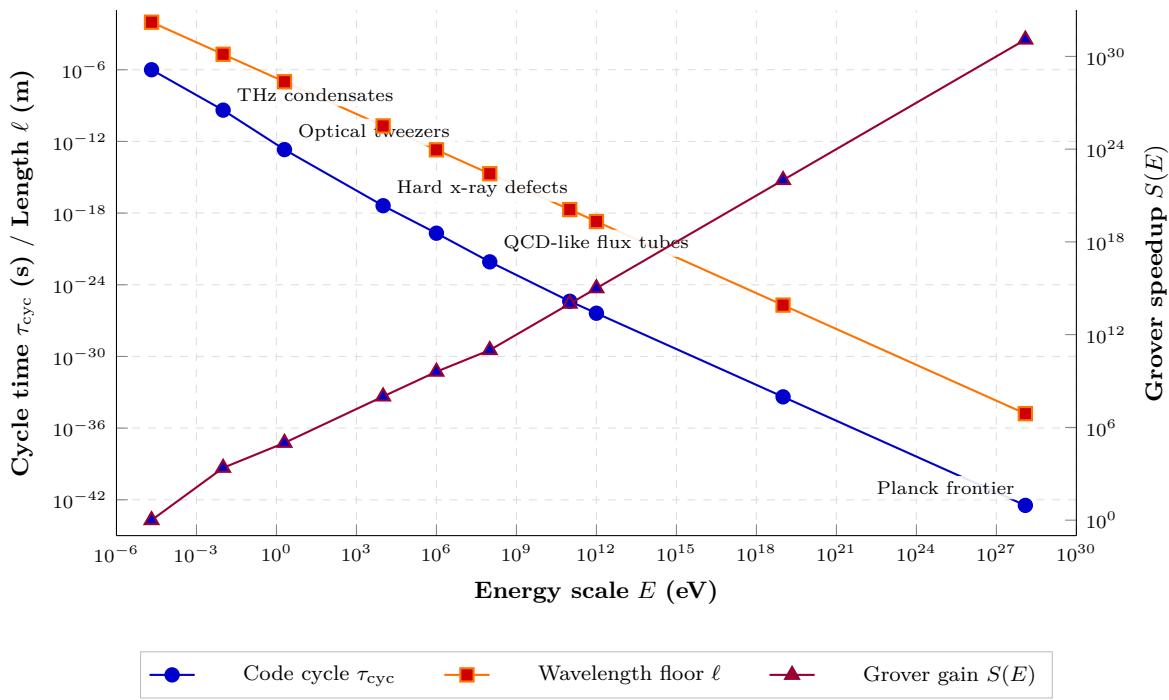
\begin{table*}[t]
  \centering
  \caption{Energy--time--length ladder for the scale-$E$ surface code. The $\kappa$ column supplies an order-of-magnitude control/measurement overhead used in Table~\ref{tab:spec-platforms}.}
  \label{tab:hesc-energy-ladder}
  \scriptsize
  \renewcommand{\arraystretch}{1.12}
  \begin{tabularx}{\linewidth}{@{}l Y Y Y Y Y@{}}
    \toprule
    Tag & $E$ [eV] & $\tau_0=h/E$ [s] & $\ell=\hbar c/E$ [m] & $\kappa(E)$ [--] & $S(E)$ [--] \\
    \midrule
    surface\_mw\_5GHz    & \num{2.067834e-05} & \num{2.000e-10} & \num{9.543e-03} & \num{5.000e+03} & \num{1.00e+00} \\
    surface\_thz\_10meV  & \num{1.000000e-02} & \num{4.136e-13} & \num{1.973e-05} & \num{1.00e+03} & \num{2.42e+03} \\
    surface\_opt\_2eV    & \num{2.000000e+00} & \num{2.068e-15} & \num{9.866e-08} & \num{1.00e+02} & \num{1.00e+05} \\
    surface\_xray\_10keV & \num{1.000000e+04} & \num{4.136e-19} & \num{1.973e-11} & \num{1.00e+01} & \num{1.00e+08} \\
    surface\_nuclear\_1MeV & \num{1.000000e+06} & \num{4.136e-21} & \num{1.973e-13} & \num{5.00e+00} & \num{4.00e+09} \\
    surface\_qcd\_100MeV & \num{1.000000e+08} & \num{4.136e-23} & \num{1.973e-15} & \num{2.00e+00} & \num{1.00e+11} \\
    surface\_ew\_100GeV  & \num{1.000000e+11} & \num{4.136e-26} & \num{1.973e-18} & \num{1.00e+00} & \num{1.00e+14} \\
    surface\_tev\_1TeV   & \num{1.000000e+12} & \num{4.136e-27} & \num{1.973e-19} & \num{1.00e+00} & \num{1.00e+15} \\
    surface\_gut\_1e16GeV& \num{1.000000e+19} & \num{4.136e-34} & \num{1.973e-26} & \num{1.00e+00} & \num{1.00e+22} \\
    surface\_planck      & \num{1.221000e+28} & \num{3.387e-43} & \num{1.616e-35} & \num{1.00e+00} & \num{1.22e+31} \\
    \bottomrule
  \end{tabularx}
  \renewcommand{\arraystretch}{1.0}
\end{table*}

\paragraph{Speculative platform analogs.}
Table~\ref{tab:spec-platforms} invents heuristic analogs---``What stands in for superconducting qubits, ion traps, or photonics at scale $E$?''---without claiming realizability. Each row keeps \emph{surface-code primitives} (local checks; rapid measurement) while swapping carriers, consistent with the scale-invariant references.~\cite{Yao1993QuantumCircuitComplexity,BernsteinVazirani1997QuantumComplexity,Kitaev2003AnyonsFTQC,Atiyah1988TQFT,Witten1988TQFT,Nayak2008NonAbelianTQC,AharonovBenOr2008Threshold}
\begin{table*}[t]
  \centering
  \caption{Speculative surface-code platforms across energy scales. Each row lists a qubit carrier analog, a stabilizer-check primitive, control/readout carriers, a dominant noise/bottleneck, and the $\kappa$ guess used in Table~\ref{tab:hesc-energy-ladder}. The analogies honor the TQFT/circuit-model equivalences of Refs.~\cite{Kitaev2003AnyonsFTQC,FreedmanKitaevLarsenWang2003TQC,Atiyah1988TQFT,Witten1988TQFT}.}
  \label{tab:spec-platforms}
  \scriptsize
  \setlength{\tabcolsep}{3pt}
  \renewcommand{\arraystretch}{1.15}
  \begin{tabularx}{\linewidth}{@{}l l Z Z Z Y@{}}
    \toprule
    Tag & $E$ [eV] & Qubit carrier analog & Stabilizer measurement primitive & Control/readout carriers & Dominant noise/bottleneck and $\kappa$ \\
    \midrule
    surface\_mw\_5GHz & $2.1{\times}10^{-5}$ & Transmon / fluxonium & Dispersive $ZZ$ via resonator; mid-circuit readout with JPA chain & Microwaves; JTWPA; cryo-CMOS & Thermal photons, leakage; $\kappa\!\sim\!5{\times}10^3$ \\
    surface\_thz\_10meV & $10^{-2}$ & Phonon / polariton qubits in piezoelectric stacks & Near-field THz exchange; piezo-mediated parity checks & THz waveguides; Schottky mixers & THz loss, detector limits; $\kappa\!\sim\!10^3$ \\
    surface\_opt\_2eV & $2$ & Rydberg or excitonic qubits in tweezer arrays & Rydberg-blockade parity checks; cavity photon counting & Narrow-line lasers; single-photon detectors & Atom loss, shot noise; $\kappa\!\sim\!10^2$ \\
    surface\_xray\_10keV & $10^{4}$ & Inner-shell excitons / M\"ossbauer resonances & Resonant x-ray scattering parity checks & XFEL-class pulses; solid-state x-ray cavities & Detector dead time, damage; $\kappa\!\sim\!10$ \\
    surface\_nuclear\_1MeV & $10^{6}$ & Nuclear-gamma dressed two-level systems & Gamma-mediated parity via coherent forward scattering & Gamma optics; pair-production calorimetry & Backgrounds, activation; $\kappa\!\sim\!5$ \\
    surface\_qcd\_100MeV & $10^{8}$ & Topological flux-tube defects in a QCD-like condensate & Parity via braiding or annihilation of flux-anyon analogs & Meson-like bosons; hadronic calorimeters & Confinement loss, hadronization noise; $\kappa\!\sim\!2$ \\
    surface\_ew\_100GeV & $10^{11}$ & Electroweak soliton / defect qubits & Parity from weak-boson scattering off defects & Effective $W/Z$ beams; neutrino heralding & Cross-section limits; $\kappa\!\sim\!1$ \\
    surface\_tev\_1TeV & $10^{12}$ & TeV-scale defect qubits in beyond-SM condensates & Local checks via short-range defect interactions & TeV photonics / phononics & Power density; $\kappa\!\sim\!1$ \\
    surface\_gut\_1e16GeV & $10^{19}$ & GUT vortex / monopole defect pairs & Syndrome from annihilation or scattering selection rules & GUT-scale gauge bosons & Catastrophic energy density; $\kappa\!\sim\!1$ \\
    surface\_planck & $1.22{\times}10^{28}$ & ``Foam-Josephson'' defects between Planck domains & Parity from gravitational Aharonov--Bohm-like phases & Planck-frequency radiative modes & Collapse limits; $\kappa\!\sim\!1$ \cite{Lloyd1999Ultimate} \\
    \bottomrule
  \end{tabularx}
  \renewcommand{\arraystretch}{1.0}
\end{table*}

\begin{figure}[p]
  \centering
  \IfFileExists{figures/high_energy_heatmap.pdf}{%
    \includegraphics[page=1,width=0.95\columnwidth]{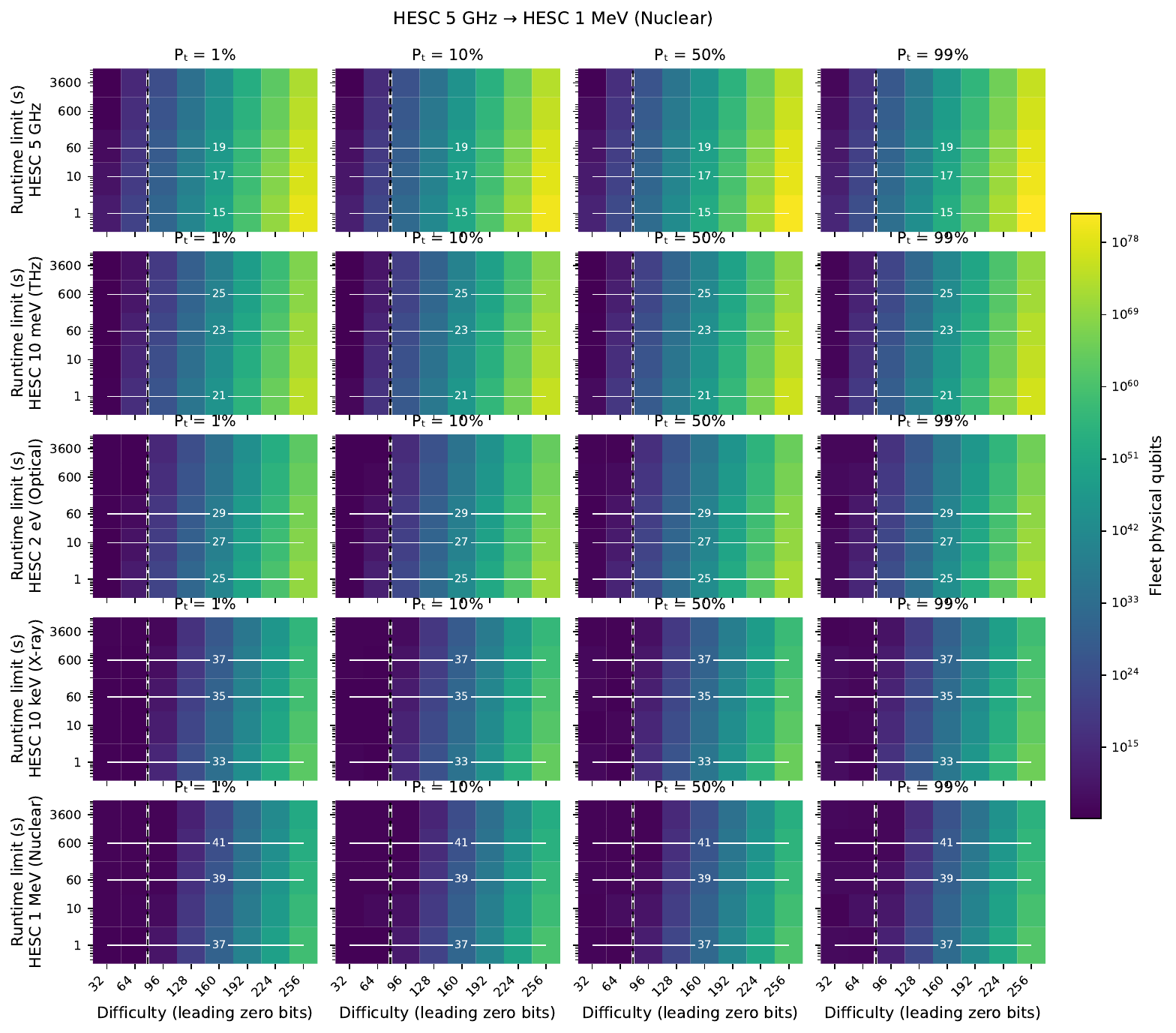}%
  }{%
    \fbox{\parbox{0.9\columnwidth}{\centering
      Figure~\ref{fig:high-energy-heatmap-low} sketches the fleet requirements implied by
      the microwave-to-MeV tiers in Table~\ref{tab:hesc-energy-ladder}.}}%
  }%
  \caption{Fleet qubits demanded by the high-energy surface-code hypothesis: microwave through nuclear tiers.
    Each row fixes an energy scale~$E$ from Table~\ref{tab:hesc-energy-ladder}, rescales the cycle time via Eq.~\eqref{eq:tau-and-ell-from-E},
    and reruns the mining sweep from Figure~\ref{fig:fleet-heatmap}.
    Advancing from microwave condensates to nuclear resonances trims at most one decade of fleet scale per tier before marked-state counts dominate.  A dashed vertical marker highlights the Bitcoin mainnet difficulty on 2025-01-01 ($b\approx78.6$)~\cite{BlockchainComDifficulty,BlockchainComChartsAPI}.}
  \label{fig:high-energy-heatmap-low}
\end{figure}

\begin{figure}[p]
  \centering
  \IfFileExists{figures/high_energy_heatmap.pdf}{%
    \includegraphics[page=2,width=0.95\columnwidth]{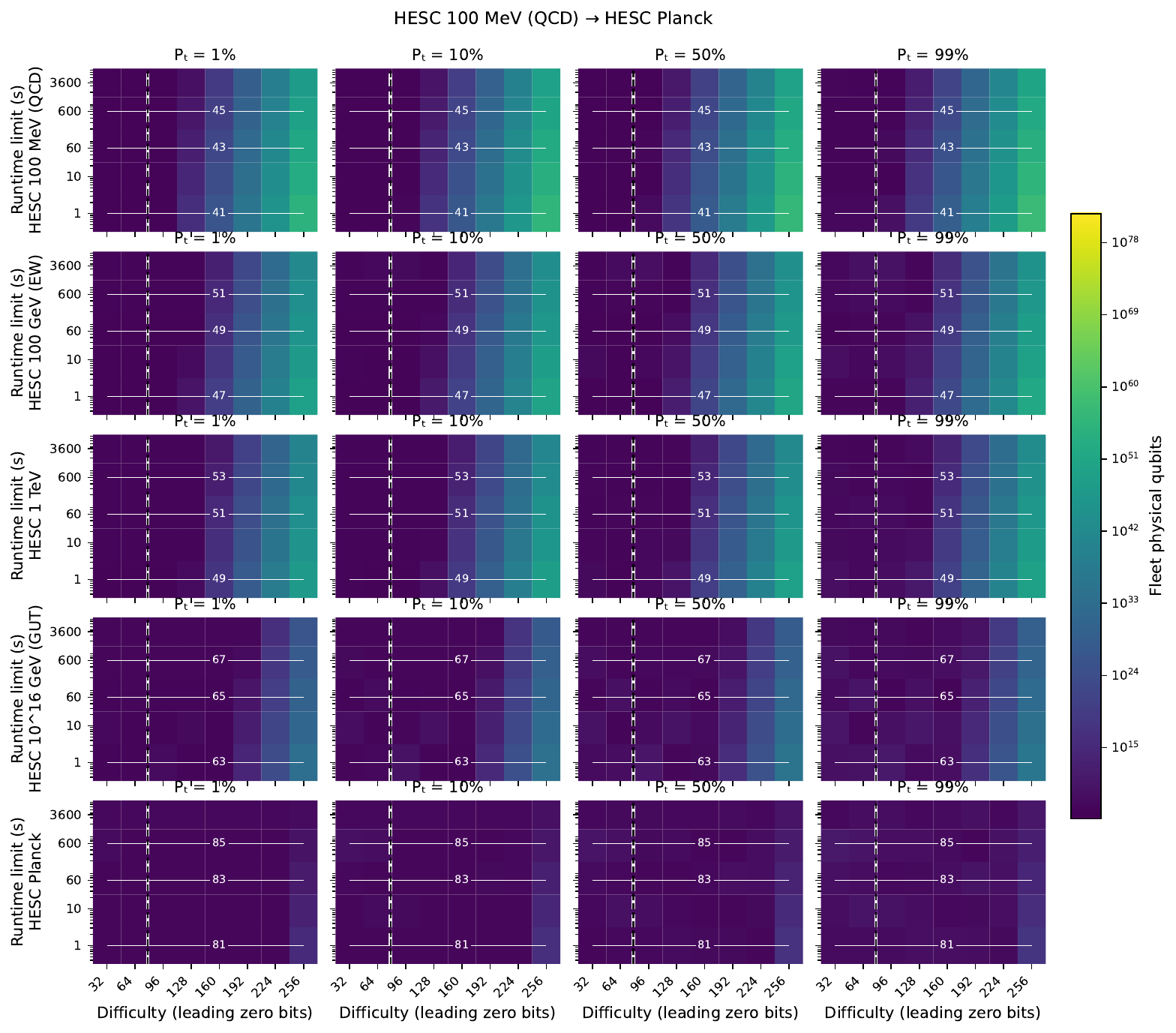}%
  }{%
    \fbox{\parbox{0.9\columnwidth}{\centering
      Figure~\ref{fig:high-energy-heatmap-high} sketches the fleet requirements implied by
      the QCD-to-Planck tiers in Table~\ref{tab:hesc-energy-ladder}.}}%
  }%
  \caption{Fleet qubits demanded by the high-energy surface-code hypothesis: QCD through Planck tiers.
    Once cycle times plunge below the target runtimes, qubit demand plateaus because fewer machines can saturate the wall-clock budget.
    Figures~\ref{fig:high-energy-heatmap-low} and~\ref{fig:high-energy-heatmap-high} together mirror the baseline Figure~\ref{fig:fleet-heatmap}, now parameterized by the HESC energy ladder.  A dashed vertical marker highlights the Bitcoin mainnet difficulty on 2025-01-01 ($b\approx78.6$)~\cite{BlockchainComDifficulty,BlockchainComChartsAPI}.}
  \label{fig:high-energy-heatmap-high}
\end{figure}

\begin{figure}[t]
  \centering
  \IfFileExists{figures/high_energy_tradeoff_curves.pdf}{%
    \includegraphics[width=0.95\columnwidth]{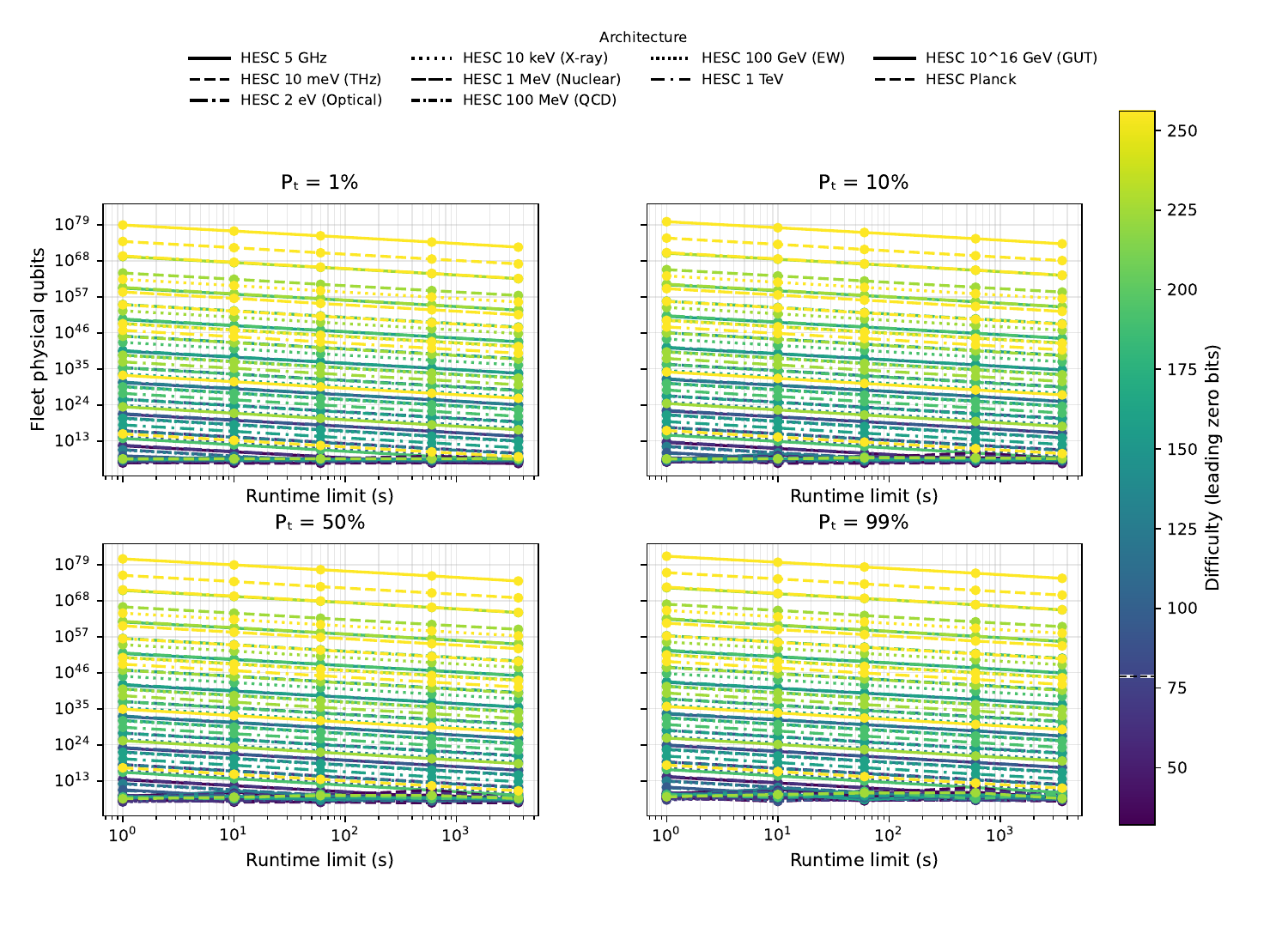}%
  }{%
    \fbox{\parbox{0.9\columnwidth}{\centering
      Figure~\ref{fig:high-energy-tradeoff} reports runtime--qubit trade-offs
      for the high-energy estimator sweep.}}
  }%
  \caption{Runtime--qubit trade-offs for the high-energy surface-code ladder.
    Each curve pairs the rescaled cycle times from Table~\ref{tab:hesc-energy-ladder}
    with the success targets of Figure~\ref{fig:fleet-tradeoff}.
    Even speculative high-energy condensates still demand orders-of-magnitude fleet growth to buy faster Grover searches.
    A dashed marker on the difficulty colorbar highlights the Bitcoin mainnet difficulty on 2025-01-01 ($b\approx78.6$)~\cite{BlockchainComDifficulty,BlockchainComChartsAPI}.}
  \label{fig:high-energy-tradeoff}
\end{figure}

\paragraph{Consequence for time-limited Grover.}
With HESC and Eq.~\eqref{eq:tau-and-ell-from-E}, the number of Grover $T$-layers that fit within a fixed wall clock scales like $S(E)$ in Table~\ref{tab:hesc-energy-ladder}. The equivalence to the circuit model~\cite{Yao1993QuantumCircuitComplexity,BernsteinVazirani1997QuantumComplexity} still holds; the microscopic carriers change while $\kappa(E)$ remains bounded.
We propagate this rescaling through the mining estimator by replacing the microwave baseline cycle time with the $\tau_{\mathrm{cyc}}(E)$ entries from Table~\ref{tab:hesc-energy-ladder}.  The resulting sweep quantifies how much faster, and therefore how much larger, a high-energy surface-code fleet must become to preserve the same Grover wall-clock bounds and success targets.  Figure~\ref{fig:high-energy-heatmap-low} marches from microwave condensates to nuclear resonances and reveals that every step toward shorter cycles trims at most one decade of fleet scale before marked-state counts dominate.  Figure~\ref{fig:high-energy-heatmap-high} leaps to QCD, electroweak, and Planck analogs; qubit demand plateaus because fewer machines can saturate the runtime budget once $\tau_{\mathrm{cyc}}$ plunges below the target runtimes.  Together with Figure~\ref{fig:high-energy-tradeoff}, these plots mirror the baseline results from \secref{sec:physical-footprint}, now parameterized by the HESC ladder, so the manuscript reports both the standard microwave assumptions and the speculative high-energy extrapolation.

\paragraph{Thermodynamic bounds at high energy scales.}
Recent full-stack power models for superconducting systems decompose qubit power into cryogenic overhead, control electronics, and gate-level dissipation~\cite{FellousAsiani2023}.
At millikelvin temperatures the Carnot penalty $T_{\mathrm{hot}}/T_{\mathrm{cold}}\sim 3\times10^{4}$ dominates, and careful cryogenic optimization can push per-qubit power down to $\sim1$--$2\,$mW---three to four orders below the empirical $\sim6\,$W extrapolations of Parker and Vermeer~\cite{ParkerVermeer2023}.
That optimistic bound depends on solid-state technological assumptions: staged dilution refrigerators, classical complementary metal--oxide--semiconductor (CMOS) control electronics at fixed temperature, and noise models calibrated to transmon-like qubits.

For the speculative high-energy platforms in Table~\ref{tab:hesc-energy-ladder}, these assumptions break down in opposite directions.
On one hand, cryogenic penalties vanish once the operating temperature approaches or exceeds room temperature (roughly $E\gtrsim 25\,$meV), removing a dominant sink in superconducting power models.
On the other hand, an intrinsic gate-power floor worsens quadratically with energy scale.
Each logical gate toggles an excitation of energy $E$, and gates arrive at rate $1/\tau_{\mathrm{cyc}}\propto E/(\kappa h)$, which yields the thermodynamic bound
\begin{equation}
  P_{\mathrm{gate}}^{\min} \;\gtrsim\; \frac{E}{\tau_{\mathrm{cyc}}} \;\sim\; \frac{E^{2}}{\kappa\, h}.
  \label{eq:gate-power-floor}
\end{equation}
At the superconducting baseline ($E\sim 20\,\mu$eV) this floor is negligible ($\sim10^{-14}\,$W), consistent with cryogenics and control electronics dominating~\cite{FellousAsiani2023}.
At nuclear scales ($E\sim1\,$MeV) the same expression returns $\sim10^{25}\,$W per qubit---stellar power for gate switching alone, before overhead.
The high-energy ladder therefore does not evade power constraints by shedding cryogenics; it trades one bottleneck for another that scales more steeply with $E$.
Read the ladder as an illustration of clock-rate ceilings rather than an engineering forecast.
Ultimate device limits remain bounded by Margolus--Levitin/Bekenstein arguments as surveyed by Lloyd~\cite{Lloyd1999Ultimate}.

\bibliographystyle{quantum}
\bibliography{kardashev}

\end{document}